\documentclass[usenatbib]{mn2e}
\usepackage{graphicx}
\usepackage{amsmath}
\usepackage{times}
\usepackage{xspace}
\usepackage{color}
\usepackage{breqn}

\usepackage[normalem]{ulem}
\usepackage{float}
\usepackage{epsfig}
\usepackage{multirow}
\usepackage{epstopdf}
\usepackage{verbatim}
\usepackage{nicefrac}
\usepackage{url}
\usepackage{subfig}
\usepackage{bm}
\usepackage[table,usenames,dvipsnames]{xcolor}
\usepackage{amssymb}
\usepackage{booktabs}
\usepackage{threeparttable}
\usepackage{amsbsy}
\usepackage{amsfonts}

\usepackage{tabularx}
\newcolumntype{R}{>{\raggedleft\arraybackslash}X}

\newcommand{\sjaddress}{\url{https://github.com/sjoudaki/kids450}\xspace}
\newcommand{\kidsaddress}{\url{http://kids.strw.leidenuniv.nl}\xspace}

\newcommand{\be}{\begin{equation}}
\newcommand{\ee}{\end{equation}}
\newcommand{\bea}{\begin{eqnarray}}
\newcommand{\eea}{\end{eqnarray}}

\newcolumntype{P}[1]{>{\centering\arraybackslash}p{#1}}

\newcommand{\halofit}{\textsc{halofit}\xspace}
\newcommand{\hmcode}{\textsc{hmcode}\xspace}
\newcommand{\theli}{\textsc{theli}\xspace}
\newcommand{\astrowise}{\textsc{Astro-WISE}\xspace}
\newcommand{\isitgr}{\textsc{ISiTGR}\xspace}
\newcommand{\gadget}{\textsc{gadget-2}\xspace}

\newcommand{\cosmomc}{\textsc{CosmoMC}\xspace}
\newcommand{\plik}{\textsc{Plik}\xspace}
\newcommand{\camb}{\textsc{CAMB}\xspace}
\newcommand{\astac}{\textsc{ASTAC}\xspace}
\newcommand{\caastro}{\textsc{CAASTRO}\xspace}

\begin{document}

\voffset=-1.5cm
\hoffset=0.45cm

\title[KiDS extended cosmologies]{KiDS-450: Testing extensions to the standard cosmological model}

\author[Joudaki et al.]{\parbox[t]{\textwidth}{Shahab Joudaki$^{1,2}$\thanks{E-mail: sjoudaki@swin.edu.au},
    Alexander Mead$^3$, Chris Blake$^1$, Ami Choi$^4$, Jelte de Jong$^5$, \\Thomas Erben$^6$, Ian Fenech Conti$^{7,8}$, Ricardo Herbonnet$^5$, Catherine Heymans$^4$, \\Hendrik Hildebrandt$^6$, Henk Hoekstra$^5$, Benjamin Joachimi$^9$, Dominik Klaes$^6$, \\Fabian K\"ohlinger$^5$, Konrad Kuijken$^5$, John McFarland$^{10}$, Lance Miller$^{11}$, \\Peter Schneider$^6$, Massimo Viola$^5$} \\ \\ $^1$ Centre for Astrophysics \&
  Supercomputing, Swinburne University of Technology, P.O.\ Box 218,
  Hawthorn, VIC 3122, Australia \\ $^2$ ARC Centre of Excellence for All-sky Astrophysics (CAASTRO) \\ $^3$ Department of Physics and Astronomy, The University of British Columbia, 6224 Agricultural Road, Vancouver, B.C., V6T 1Z1, Canada \\ $^4$ Scottish Universities Physics Alliance, 
    Institute for Astronomy, University of Edinburgh, Royal Observatory, Blackford
  Hill, Edinburgh, EH9 3HJ, U.K. \\ $^5$ Leiden Observatory, Leiden University, Niels Bohrweg 2, 2333 CA Leiden, the Netherlands \\ $^6$ Argelander Institute for Astronomy, University of Bonn, Auf dem Hugel 71, 53121 Bonn, Germany \\ 
  $^7$ Institute of Space Sciences and Astronomy (ISSA), University of Malta, Msida MSD 2080 \\ 
  $^8$ Department of Physics, University of Malta, Msida, MSD 2080, Malta \\  
  $^9$ Department of Physics and Astronomy, University College London, London WC1E 6BT, U.K. \\ $^{10}$ Kapteyn Astronomical Institute, P.O. Box 800, 9700 AV Groningen, the Netherlands \\ $^{11}$ Department of Physics, University of Oxford, Denys Wilkinson Building, Keble Road, Oxford OX1 3RH, U.K.}

\pubyear{2017}
\date{\today}

\maketitle

\begin{abstract}
We test extensions to the standard cosmological model with weak gravitational lensing tomography using 450~deg$^2$ of imaging data from the Kilo Degree Survey (KiDS). In these extended cosmologies, which include massive neutrinos, nonzero curvature, evolving dark energy, modified gravity, and running of the scalar spectral index, we also examine the discordance between KiDS and cosmic microwave background measurements from Planck. The discordance between the two datasets is largely unaffected by a more conservative treatment of the lensing systematics and the removal of angular scales most sensitive to nonlinear physics. The only extended cosmology that simultaneously alleviates the discordance with Planck and is at least moderately favored by the data includes evolving dark energy with a time-dependent equation of state (in the form of the $w_0 - w_a$ parameterization). In this model, the respective $S_8 = \sigma_8 \sqrt{\Omega_{\mathrm m}/0.3}$ constraints agree at the $1\sigma$ level, and there is `substantial concordance' between the KiDS and Planck datasets when accounting for the full parameter space. Moreover, the Planck constraint on the Hubble constant is wider than in $\Lambda$CDM and in agreement with the \citet{riess16} direct measurement of $H_0$. The dark energy model is moderately favored as compared to $\Lambda$CDM when combining the KiDS and Planck measurements, and remains moderately favored after including an informative prior on the Hubble constant. In both of these scenarios, marginalized constraints in the $w_0-w_a$ plane are discrepant with a cosmological constant at the $3\sigma$ level. Moreover, KiDS constrains the sum of neutrino masses to 4.0 eV (95\% CL), finds no preference for time or scale dependent modifications to the metric potentials, and is consistent with flatness and no running of the spectral index.
The analysis code is publicly available at \sjaddress.
\end{abstract}


\section{Introduction}
\label{Introduction}
\setcounter{footnote}{0}
\renewcommand{\thefootnote}{\arabic{footnote}}

The weak gravitational lensing measurements of the Kilo Degree Survey (KiDS; \citealt{dejong13, kuijken15, Hildebrandt16, fc16}) and cosmic microwave background measurements of the Planck satellite \citep{planck15,planck15like} have been found to be substantially discordant \citep{Hildebrandt16}. 
When quantifying this discordance in terms of the $S_8 = \sigma_8 \sqrt{\Omega_{\mathrm m}/0.3}$ parameter combination that KiDS measures well (as the amplitude of the lensing measurements roughly scale as $S_8^{2.5}$; \citealt{js97}), we find a discordance at the level of $2.3\sigma$~\citep{Hildebrandt16}.

While the area of systematic uncertainties in weak lensing will continue to improve with future studies of KiDS, this discordance is seemingly not resolved
even after accounting for intrinsic galaxy alignments, baryonic effects in the nonlinear matter power spectrum, and photometric redshift uncertainties, along with additive and multiplicative shear calibration corrections \citep{Hildebrandt16}. Assuming Planck itself is not suffering from an unknown systematic (e.g.~\citealt{addison15, planckinterm}), we are therefore motivated to examine to what degree this discordance can be alleviated by an extension to the fiducial treatment of the lensing systematics and by an expansion of the standard cosmological constant + cold dark matter ($\Lambda$CDM) model.

Beyond our fiducial treatment of the lensing systematics, which is identical to the approach in \citet{Hildebrandt16}, we consider the impact of a possible redshift dependence in the modeling of the intrinsic galaxy alignments, along with wider priors on the intrinsic alignment amplitude and baryon feedback affecting the nonlinear matter power spectrum. We do not consider introducing any free parameters in the modeling of the photometric redshift uncertainties, but instead continue to sample over a large range of bootstrap realizations from our `weighted direct calibration' (DIR) method that encapsulate the uncertainty in the redshift distribution. Separately, we also examine the discordance between KiDS and Planck when taking the conservative approach of discarding all angular bins in the KiDS measurements that are sensitive to nonlinear physics.

In addition to the lensing systematics,
the cosmological extensions that we consider are active neutrino masses, nonzero curvature, evolving dark energy (both with a constant equation of state, and with a time-dependent parameterization), modifications to gravity (by modifying the Poisson equation and deflection of light), and nonzero running of the scalar spectral index. We take a conservative approach and consider these extensions independently, but also consider a case where curvature and evolving dark energy are analyzed jointly. In our Markov Chain Monte Carlo (MCMC) analyses, we vary the new degrees of freedom of the extended cosmological models along with the standard $\Lambda$CDM and lensing systematics parameters (and CMB degrees of freedom when applicable). We list the priors associated with these degrees of freedom in Table~\ref{table:priors}.

\begin{table}
\begin{center}
\caption{Priors on the cosmological and lensing systematics parameters.
The cosmological parameters in the first third of this table are defined as `vanilla' parameters, and $\theta_{\mathrm{s}}$ denotes the angular size of the sound horizon at the redshift of last scattering. 
We always vary the vanilla parameters and lensing systematics parameters (IA and baryon feedback amplitudes) in our MCMC calculations. Following \citet{Hildebrandt16}, we also always account for photometric redshift uncertainties by using 1000 bootstrap realizations of the tomographic redshift distributions (see Section~\ref{theobs}).
We emphasize that the Hubble constant is a derived parameter.
Unlike the analysis in \citet{Hildebrandt16}, we fiducially do not impose an informative prior on the Hubble constant from \citet{riess16}, and we impose a weaker informative prior on the baryon density, as described in Section~\ref{theobs}. When we do impose an informative prior on the Hubble constant in specific instances, this is manifested as a uniform $\pm5\sigma$ prior from \citet{riess16}, such that $0.64 < h < 0.82$.
The optical depth is only varied when the CMB is considered. The extended cosmological parameters 
are varied as described in Sections~\ref{neum} to \ref{runningsec}. 
}
\begin{tabular}{lll}
\toprule
Parameter & Symbol & Prior\\
\midrule
Cold dark matter density & $\Omega_{\mathrm c}h^2$ & $[0.001, 0.99]$\\
Baryon density & $\Omega_{\mathrm b}h^2$ & $[0.013, 0.033]$\\
100 $\times$ approximation to $\theta_{\mathrm s}$ & $100 \theta_{\rm MC}$ & $[0.5, 10]$\\
Amplitude of scalar spectrum & $\ln{(10^{10} A_{\mathrm s})}$ & $[1.7, 5.0]$\\
Scalar spectral index & $n_{\mathrm s}$ & $[0.7, 1.3]$\\
Optical depth & $\tau$ & $[0.01, 0.8]$ \\
Dimensionless Hubble constant & $h$ & $[0.4, 1.0]$ \\
Pivot scale $[{\rm{Mpc}}^{-1}]$ & $k_{\rm pivot}$ & 0.05 \\
\midrule
IA amplitude & $A_{\rm IA}$ & $[-6, 6]$\\
{\it~~~--~extended case} &  & $[-20, 20]$\\
IA redshift dependence & $\eta_{\rm IA}$ & [0, 0]\\
{\it~~~--~extended case} &  & $[-20, 20]$\\
Feedback amplitude & $B$ & $[2, 4]$\\
{\it~~~--~extended case} &  & $[1, 10]$\\
\midrule
MG bins (modifying grav. const.) & $Q_i$ & $[0, 10]$\\
MG bins (modifying deflect. light) & $\Sigma_j$ & $[0, 10]$\\
Sum of neutrino masses [eV] & $\sum m_\nu$ & $[0.06, 10]$\\
Effective number of neutrinos & $N_{\rm eff}$ & [1.046, 10]\\
Constant dark energy EOS & $w$ & $[-3, 0]$\\
Present dark energy EOS & $w_0$ & $[-3, 0]$\\
Derivative of dark energy EOS & $w_a$ & $[-5, 5]$\\
Curvature & $\Omega_k$ & $[-0.15, 0.15]$\\
Running of the spectral index & ${{\mathrm d}n_{\mathrm s} / {\mathrm d}\ln k}$ & $[-0.5, 0.5]$\\
\bottomrule
\end{tabular}
\label{table:priors}
\end{center}
\end{table}

Beyond the KiDS-Planck discordance, earlier lensing observations by the Canada-France-Hawaii Telescope Lensing Survey (CFHTLenS; \citealt{heymans12,hildebrandt12,erben13,miller13}) were also found to exhibit a similar tension with Planck (e.g.~\citealt{planck13, maccrann15, planck15, kohlinger15, joudaki16}). 
This CFHTLenS-Planck discordance has been explored in the context of extensions to the standard $\Lambda$CDM model and systematic uncertainties in the lensing measurements (e.g.~\citealt{maccrann15, kohlinger15, kns15, leistedt14,battye15,enqvist15,dVMS16,dipd15,joudaki16, liu16, alsing16}).
Meanwhile, lensing observations by the Deep Lens Survey (DLS, \citealt{jee2016}) exhibit a mild discrepancy with KiDS (at $\sim1.5\sigma$ in $S_8$), and observations by the Dark Energy Survey (DES, \citealt{dessv}) have sufficiently large uncertainties that they agree both with CFHTLenS/KiDS and Planck.

\begin{figure*}
\includegraphics[width=0.95\hsize]{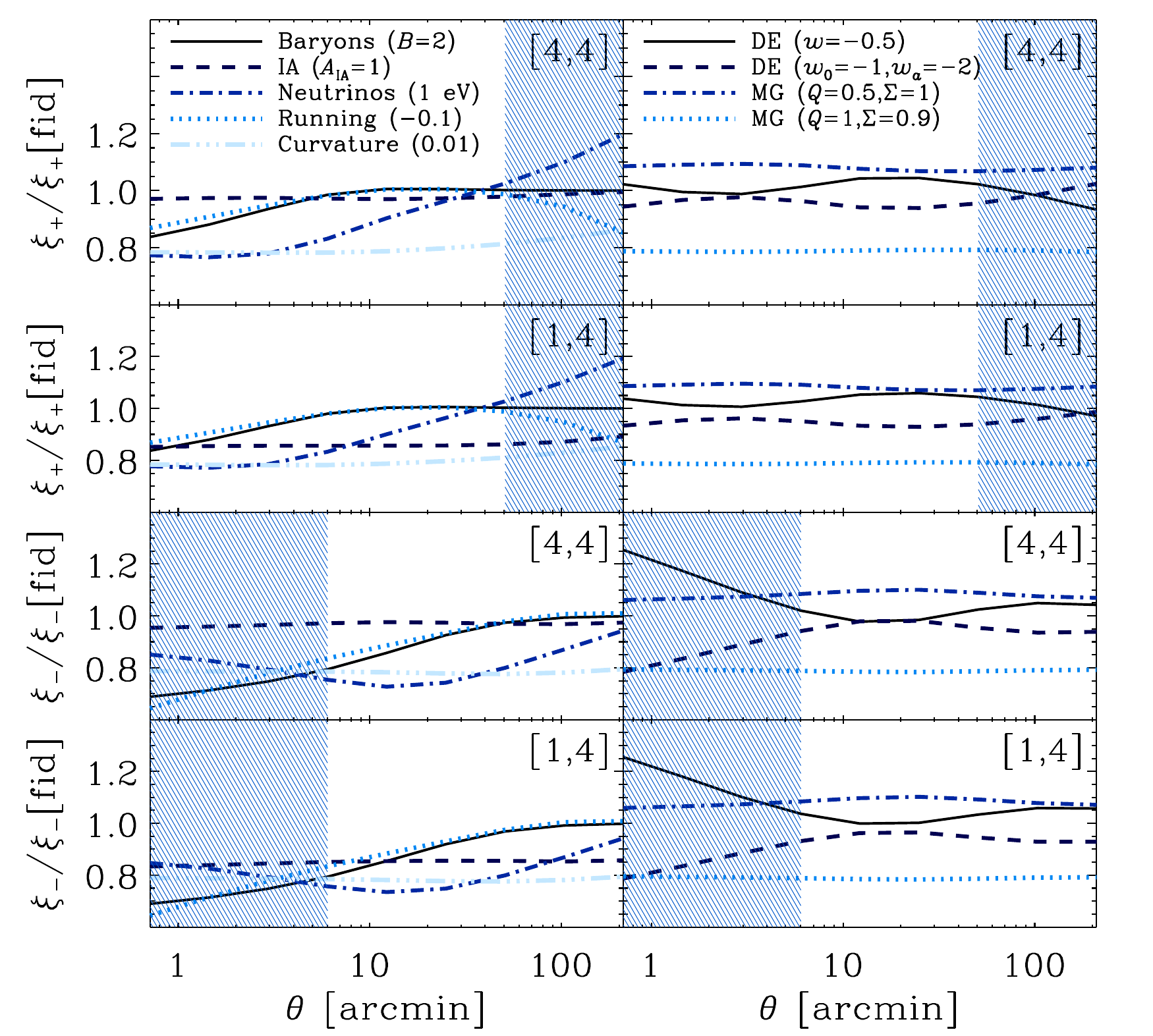}
\vspace{-0.3em}
\caption{\label{figcorr} Ratio of shear correlation functions $\xi_{\pm}^{ij}(\theta)$ for tomographic bin combinations $\{1,4\}$ and $\{4,4\}$, taken for each extended parameter with respect to a flat $\Lambda$CDM model including no systematic uncertainties (denoted as $\xi_{\pm}[{\rm{fid}}]$). Parameter definitions are listed in Table~\ref{table:priors}. For each perturbation, we keep all primary parameters fixed. These primary parameters include $\{\Omega_{\mathrm c}h^2, \Omega_{\mathrm b}h^2, \theta_{\mathrm{MC}}, \ln{(10^{10} A_{\mathrm s})}, n_{\mathrm s}\}$, 
along with the intrinsic alignment amplitude $A_{\rm IA}$ and baryon feedback amplitude $B$ when not explicitly varied (but not for instance the Hubble constant as it is a derived parameter). 
The curvature case corresponds to $\Omega_k = 0.01$, the neutrino mass case corresponds to $\sum m_{\nu} = 1~{\rm eV}$, and the case with nonzero running corresponds to ${{\mathrm d}n_{\mathrm s} / {\mathrm d}\ln k} = -0.1$. The modified gravity parameters Q and $\Sigma$ modify the gravitational constant and deflection of light, respectively. The dark energy equation of state can either be constant~($w$), or possess a time-dependence with $w_0$ and $w_a$. The shaded regions correspond to angular scales that are masked out in the KiDS analysis.}
\end{figure*}

As we focus on the discordance between KiDS and Planck in the context of extended cosmologies, we also examine whether these cosmologies can simultaneously resolve the approximately $3\sigma$ tension between Planck and local measurements of the Hubble constant based on the cosmic distance ladder \citep{riess11, riess16}. In particular, it has been suggested that the tension in the Hubble constant can be resolved by invoking non-standard physics in the dark energy and dark radiation sectors (most recently, e.g.~\citealt{bernal16,dVMS16,divalentino16,grandis16,kk16,archidiacono16,riess16}). 

Beyond questions of dataset concordance, we examine to what extent the additional degrees of freedom in the extended cosmological models are constrained by the data (when KiDS and Planck are not in tension), and to what degree the extended models are favored by the data from the point of view of model selection, using statistical tools such as the deviance information criterion (DIC). In assessing the viability of the extended cosmologies, it is not sufficient that they alleviate the discordance with Planck, but they need to be favored by the data from the point of model selection as compared to the standard cosmology.

In Section~\ref{Methodology}, we describe the KiDS measurements and underlying statistics used to analyze them. In Section~\ref{Results}, we constrain extensions to the fiducial treatment of the lensing systematics and to the standard cosmological model, in the form of massive neutrinos, curvature, evolving dark energy, modified gravity, and running of the scalar spectral index. We examine to what degree the extended cosmologies are favored by KiDS and Planck, and to what extent they help to alleviate the $\Lambda$CDM discordance between the KiDS and Planck datasets. In Section~\ref{conclusions}, we conclude with a discussion of our results.

\begin{table}
\begin{center}
\caption{Exploring changes in $\chi^2_{\rm eff}$ and DIC for different extensions to the standard cosmological model (given the priors in Table~\ref{table:priors}, lensing systematics always included).
The reference $\Lambda$CDM model (with fiducial treatment of lensing systematics) gives $\chi^2_{\rm eff} = 162.3$ and $\rm{DIC} = 177.4$ for KiDS (marginally different from the values in \citealt{Hildebrandt16} due to wider priors on the baryon density and Hubble constant), $\chi^2_{\rm eff} = 11265.4$ and $\rm{DIC} = 11297.5$ for Planck (marginal change from \citealt{planck15} due to different priors), $\chi^2_{\rm eff} = 11438.6$ and $\rm{DIC} = 11477.8$ for the joint analysis of KiDS and Planck, $\chi^2_{\rm eff} = 11439.0$ and $\rm{DIC} = 11478.0$ for the joint analysis of KiDS and Planck with an informative Hubble constant prior from \citet{riess16}. 
Negative values indicate preference in favor of the extended model as compared to fiducial $\Lambda$CDM.
}
\begin{tabular}{p{4.3cm}>{\raggedleft}p{1.3cm}>{\raggedleft\arraybackslash}p{1.4cm}}
\toprule
Model & $\Delta\chi^2_{\rm eff}$ & $\Delta{\rm DIC}$\\
\midrule
$\Lambda$CDM (extended systematics) &  & \\
{\it~~~--~KiDS} & $-2.1$ & $2.4$\\
{\it~~~--~Planck} & $0$ & $0$\\
{\it~~~--~KiDS+Planck} & $-0.87$ & $2.7$\\
Neutrino mass &  & \\
{\it~~~--~KiDS} & $0.10$ & $2.7$\\
{\it~~~--~Planck} & $2.0$ & $3.4$\\
{\it~~~--~KiDS+Planck} & $2.9$ & $3.3$\\
Curvature &  & \\
{\it~~~--~KiDS} & $-0.96$ & $-0.22$\\
{\it~~~--~Planck} & $-5.8$ & $-4.3$\\
{\it~~~--~KiDS+Planck} & $-0.22$ & $0.31$\\
Dark energy (constant $w$) &  & \\
{\it~~~--~KiDS} & $0.074$ & $2.3$\\
{\it~~~--~Planck} & $-3.1$ & $-0.20$\\
{\it~~~--~KiDS+Planck} & $-5.5$ & $-5.4$\\
{\it~~~--~KiDS+Planck+$H_0$} & $-3.4$ & $-2.9$\\
Dark energy ($w_0-w_a$) &  & \\
{\it~~~--~KiDS} & $-0.35$ & $0.95$\\
{\it~~~--~Planck} & $-3.2$ & $-1.1$\\
{\it~~~--~KiDS+Planck} & $-6.4$ & $-6.8$\\
{\it~~~--~KiDS+Planck+$H_0$} & $-6.5$ & $-6.4$\\
Curvature + dark energy (constant $w$) &  & \\
{\it~~~--~KiDS} & $-0.44$ & $0.30$\\
{\it~~~--~Planck} & $-6.2$ & $-3.7$\\
{\it~~~--~KiDS+Planck} & $-5.8$ & $-3.6$\\
{\it~~~--~KiDS+Planck+$H_0$} & $-3.6$ & $-2.0$\\
Modified gravity (fiducial scales) &  & \\
{\it~~~--~KiDS} & $-3.6$ & $-0.094$\\
{\it~~~--~Planck} & $-4.0$ & $5.7$\\
{\it~~~--~KiDS+Planck} & $-4.2$ & $0.77$\\
Modified gravity (large scales) &  & \\
{\it~~~--~KiDS} & $-6.4$ & $5.9$\\
{\it~~~--~Planck} & $-4.0$ & $5.7$\\
{\it~~~--~KiDS+Planck} & $-6.5$ & $2.4$\\
Running of the spectral index &  & \\
{\it~~~--~KiDS} & $-1.1$ & $0.27$\\
{\it~~~--~Planck} & $-0.058$ & $0.68$\\
{\it~~~--~KiDS+Planck} & $0.46$ & $1.1$\\
\bottomrule
\end{tabular}
\label{table:chidic}
\end{center}
\end{table}

\begin{table}
\begin{center}
\caption{Assessing the level of concordance between KiDS and Planck as quantified by $T(S_8)$ defined in equation~(\ref{eqn:ts8}), and $\log \mathcal{I}$ (base 10) defined in equation~(\ref{eqn:logi1}). The $\Lambda$CDM results with fiducial treatment of the systematic uncertainties differ marginally from \citet{Hildebrandt16} due to our wider priors on the Hubble constant and baryon density.
}
\begin{tabular}{p{4.3cm}>{\raggedleft}p{1.3cm}>{\raggedleft\arraybackslash}p{1.4cm}}
\toprule
Model & $T(S_8)$ & $\log \mathcal{I}$\\
\midrule
$\Lambda$CDM &  & \\
{~~---~~fiducial systematics} & 2.1$\sigma$ & -0.63\\
{~~---~~extended systematics} & 1.8$\sigma$ & -0.70\\
{~~---~~large scales} & 1.9$\sigma$ & -0.62\\
Neutrino mass & 2.4$\sigma$  & -0.011 \\
Curvature & 3.5$\sigma$ & -1.7 \\
Dark energy (constant $w$) & 0.89$\sigma$ & 0.99 \\
Dark energy ($w_0-w_a$) & 0.91$\sigma$ & 0.82 \\
Curvature + dark energy (constant $w$) & 2.5$\sigma$ & -0.59 \\
Modified gravity (fiducial scales) & 0.49$\sigma$ & 0.42 \\
Modified gravity (large scales) & 0.83$\sigma$ & 1.4 \\
Running of the spectral index & 2.3$\sigma$ & -0.66 \\
\bottomrule
\end{tabular}
\label{table:conc}
\end{center}
\end{table}

\section{Methodology}
\label{Methodology}

We give a description of the KiDS and Planck datasets used and computational approach in Section~\ref{theobs}, our statistical analysis tools in Section~\ref{modsec}, and baseline configurations in Section~\ref{basesec}.

\subsection{Theory and measurements}
\label{theobs}

We follow the approach presented in \citet{Hildebrandt16} to compute the weak lensing theory and associated systematic uncertainties, using the same KiDS-450 cosmic shear tomography measurements, redshift distributions, analytic covariance matrix, and cosmology fitting pipeline.

The lensing observables are given by the 2-point shear correlation functions $\xi_{\pm}^{ij}(\theta)$, for tomographic bin combination $\{i,j\}$ at angle $\theta$ (e.g.~see equations~2~to~5 in \citealt{Hildebrandt16}). 
The KiDS-450 dataset \citep{kuijken15, Hildebrandt16, fc16} covers an effective area of 360~deg$^2$, with a median redshift of $z_{\mathrm m} = 0.53$, and an effective number density of $n_{\rm eff} = 8.5$ galaxies arcmin$^{-2}$. The raw pixel data is processed by \theli \citep{erben13} and \astrowise \citep{begeman13,dejong15}, while the shears are measured using ${\it{lens}}$fit \citep{miller13}.
The dataset consists of 4 tomographic bins between $z_{\mathrm B} = 0.1$ to $z_{\mathrm B} = 0.9$ (equal widths $\Delta z_{\mathrm B} = 0.2$), where $z_{\mathrm B}$ is the best-fitting redshift output by BPZ \citep{benitez2000}. For each tomographic bin, the measurements cover 7 angular bins between 0.5 to 72 arcmins in $\xi_{+}^{ij}(\theta)$ and 6 angular bins logarithmically spaced between 4.2 to 300 arcmins in $\xi_{-}^{ij}(\theta)$. In other words, considering 9 angular bins with central values at $[0.713, 1.45, 2.96, 6.01, 12.2, 24.9, 50.7, 103, 210 ]$ arcmins, the last two angular bins are masked out for $\xi_{+}^{ij}(\theta)$ and the first three angular bins are masked out for $\xi_{-}^{ij}(\theta)$. This equates to a total of 130 elements in our data vector. We use an analytical model that accounts for both Gaussian and non-Gaussian contributions in calculating the covariance matrix of our data, as described in \citet[further see Joachimi et al., in prep.]{Hildebrandt16}.

Given external overlapping spectroscopic surveys, we calibrate the photometric redshift distributions using the `weighted direct calibration' (DIR) method in \citet{Hildebrandt16}, with uncertainties and correlations between tomographic bins obtained from 1000 bootstrap realizations (using each bootstrap sample for a fixed number of MCMC iterations). We account for intrinsic galaxy alignments, given by correlations of intrinsic ellipticities of galaxies with each other and with the shear of background sources, by varying an unknown amplitude $A_{\rm IA}$ and redshift dependence $\eta_{\rm IA}$ (e.g.~see equations~4~to~7 in \citealt{joudaki16}). 
As a result, the `shear-intrinsic' and `intrinsic-intrinsic' power spectra are proportional to 
$A_{\rm IA} (1+z)^{\eta_{\rm IA}}$ and $A_{\rm IA}^2 (1+z)^{2\eta_{\rm IA}}$, respectively.
Since the mean luminosity is effectively the same across tomographic bins in KiDS, we do not consider a possible luminosity dependence of the intrinsic alignment signal \citep{Hildebrandt16}. The standard power-law extension for redshift and luminosity were introduced to account for their dependence in the coupling between galaxy shape and tidal field, which is unconstrained in any IA model. A weakness of this extension is that it is purely empirical, but it has been fit to data and demonstrated to work well (e.g.~\citealt{Joachimi11}). We also do not account for a scale dependence as there is currently no indication for it from data.

We include baryonic effects in the nonlinear matter power spectrum with \hmcode (\citealt{Mead15, Mead16}, now incorporated in \camb; \citealt{LCL}), which is a new accurate halo model calibrated to the Coyote dark matter simulations (\citealt{Coyote4}, references therein) and the OverWhelmingly Large (OWL) hydrodynamical simulations \citep{Schaye10,Daalen11}.
In \hmcode, the feedback amplitude $B$ is a free parameter that is varied in our analysis.
In this one-parameter baryon model, $B$ modifies the halo mass-concentration relation and simultaneously lightly changes the overall shape of the halo density profile in a way that accounts for the main effects of baryonic feedback in the nonlinear matter power spectrum \citep{Mead15}.

The impact of these systematic uncertainties are included in the \cosmomc \citep{Lewis:2002ah} fitting pipeline used in \citet{Hildebrandt16}, first presented in \citet{joudaki16}. Fiducially, we use the same priors on the parameters $A_{\rm IA}$, $\eta_{\rm IA}$, and $B$ as in \citet{Hildebrandt16}, listed in Table~\ref{table:priors}. 
We do not include additional degrees of freedom in our analyses for the additive and multiplicative shear calibration corrections \citep{fc16}, but incorporate these directly in our data \citep{Hildebrandt16}. Our setup agrees with the fiducial setup of systematic uncertainties in \citet{Hildebrandt16}, given by the `KiDS-450' row in their Table 4.

Our parameter priors are identical to the priors given in \citet{Hildebrandt16}, with the exception of the baryon density and Hubble constant. We impose the conservative prior $0.013 < \Omega_{\mathrm b} h^2 < 0.033$ on the baryon density (motivated by the BBN constraints in \citealt{burles01,pdg,cyburt16}) and $0.4 < h < 1.0$ on the dimensionless Hubble constant (which is a derived parameter). These choices can be contrasted with the tighter $0.019 < \Omega_{\mathrm b} h^2 < 0.026$ and $0.64 < h < 0.82$ priors in \citealt{Hildebrandt16}. 
The uniform Hubble constant prior in \citet{Hildebrandt16} encapsulates the $\pm5\sigma$ range from the direct measurement of \citet{riess16}, where $h = 0.732 \pm 0.017$, and extends beyond the Planck CMB constraint on this parameter~(\citealt{planck15}, where $h = 0.673 \pm 0.010$ for TT+lowP).
Our prior choices are more conservative than in \citet{Hildebrandt16} because they may otherwise have a significant impact on the extended cosmology constraints (unlike e.g. $S_8$ in $\Lambda$CDM which is robust to both choices of priors). However, we do consider specific cases where the \citet{riess16} prior on the Hubble constant is employed (e.g. see the dark energy results in Table~\ref{table:chidic}).

In addition to examining extensions to the standard cosmological model with the KiDS-450 dataset, and assessing their significance from a model selection standpoint, we consider the impact of these extensions on the discordance between KiDS and Planck (reported in \citealt{Hildebrandt16}). To this end, the Planck measurements \citep{planck15,planck15like} that we use are the CMB temperature and polarization on large angular scales, limited to multipoles $\ell \leq 29$ (i.e.~low-$\ell$ TEB likelihood), and the CMB temperature on smaller angular scales (via the \plik TT likelihood). Thus, we conservatively do not consider Planck polarization measurements on smaller angular scales (which would increase the discordance with KiDS slightly), and we also do not consider Planck CMB lensing measurements (which would decrease the discordance with KiDS slightly).

\subsection{Model selection and dataset concordance}
\label{modsec}

As we consider extensions to the standard cosmological model, we mainly aim to address two questions. The first question pertains to model selection, i.e. whether the extended model is favored as compared to $\Lambda$CDM. To aid in this aim, we follow \citet{joudaki16} in using the Deviance Information Criterion (DIC;~\citealt{spiegelhalter02}, also see \citealt{ktp06}, \citealt{liddle07}, \citealt{trotta08}, and \citealt{spiegelhalter14}), given by the sum of two terms:
\begin{equation}
{\rm{DIC}} \equiv {\chi^2_{\rm eff}(\hat{\theta})} + 2p_D. 
\label{dicdef}
\end{equation}
Here, the first term consists of the best-fit effective ${\chi^2_{\rm eff}(\hat{\theta})}  = -2 \ln {\mathcal{L}}_{{\rm max}}$, where ${\mathcal{L}}_{\rm max}$ is the maximum likelihood of the data given the model, and $\hat{\theta}$ is the vector of varied parameters at the maximum likelihood point. The second term is the `Bayesian complexity,' $p_D = \overline{\chi^2_{\rm eff}(\theta)} - {\chi^2_{\rm eff}(\hat{\theta})}$, where the bar denotes the mean over the posterior distribution.
Thus, the DIC is composed of the sum of the goodness of fit of a given model and its Bayesian complexity, which is a measure of the effective number of parameters, and acts to penalize more complex models.
For reference, a difference in $\chi^2_{\rm eff}$ of 10 between two models corresponds to a probability ratio of 1 in 148, and we therefore take a positive difference in DIC of 10 to correspond to strong preference in favor of the reference model ($\Lambda$CDM), while an equally negative DIC difference corresponds to strong preference in favor of the extended model. We take $\Delta{\rm DIC} = 5$ to constitute moderate preference in favor of the model with the lower DIC estimate, while differences close to zero do not particularly favor one model over the other.

In \citet{Hildebrandt16}, we found that the cosmological constraints from the KiDS-450 dataset are overall internally consistent, i.e. the constraints agree despite a range of changes to the treatment of the systematic uncertainties (e.g.~see Figure 10 therein). The cosmological constraints from KiDS also agree with previous lensing analyses from CFHTLenS (see \citealt{joudaki16} and references therein) and the Dark Energy Survey~\citep{dessv}, along with pre-Planck CMB measurements from WMAP9, ACT, and SPT \citep{calabrese13}. However, KiDS does disagree with Planck \citep{planck15} at the $2\sigma$ level in $S_8 = \sigma_8 \sqrt{\Omega_{\mathrm m}/0.3}$, and this tension can seemingly not be resolved by the systematic uncertainties \citep{Hildebrandt16}.

The second question that we aim to address therefore pertains to whether an extension to the fiducial treatment of the lensing systematic uncertainties or the standard cosmological model can alleviate or completely remove the tension between KiDS and Planck. Since current lensing data mainly constrain the $S_8$ parameter combination well, we quantify the tension $T$ in this parameter, via
\begin{equation}
T(S_8) = \left|\overline{S_8^{D_1}} - \overline{S_8^{D_2}}\right| / \sqrt{\sigma^2\left(S_8^{D_1}\right) + \sigma^2\left(S_8^{D_2}\right)} ,
\label{eqn:ts8}
\end{equation}
where the datasets $D_1$ and $D_2$ refer to KiDS and Planck, respectively, the vertical bars extract the absolute value of the encased terms, the horizontal bars again denote the mean over the posterior distribution, and $\sigma$ refers to the symmetric 68\% confidence interval about the mean.

Moreover, to better capture the overall level of concordance or discordance between the two datasets, we calculate a diagnostic grounded in the DIC \citep{joudaki16}:
\begin{equation}
{\mathcal{I}}(D_1, D_2) \equiv \exp\{{-{\mathcal{G}}(D_1, D_2)/2}\}, 
\label{eqn:logi1}
\end{equation}
such that
\begin{equation}
{\mathcal{G}}(D_1, D_2) = {{{\rm{DIC}}(D_1 \cup D_2)} - {{\rm{DIC}}(D_1) - {{\rm{DIC}}(D_2)}}},
\label{eqn:logi2}
\end{equation}
where ${{\rm{DIC}}(D_1 \cup D_2)}$ is obtained from the combined analysis of the datasets. 
Thus, $\log \mathcal{I}$ is positive when two datasets are in concordance, and negative when the datasets are discordant, with values following Jeffreys' scale (\citealt{jeffreys}, \citealt{kr95}), such that $\log \mathcal{I}$ in excess of $\pm 1/2$ is considered `substantial', in excess of $\pm 1$ is considered `strong', and in excess of $\pm 2$ is considered `decisive' (corresponding to a probability ratio in excess of 100). 
In \citet{joudaki16}, this concordance test was found to largely agree with the analogous diagnostic based on the Bayesian evidence (e.g.~\citealt{mrs06, raveri15}), and enjoys the benefit of being more readily obtained from existing MCMC chains.
Our particular approach for propagating photometric redshift uncertainties into the analysis moreover makes the calculation of the evidence non-trivial.

\subsection{Baseline settings}
\label{basesec}

Our cosmology analysis is enabled by a series of MCMC runs, using the \cosmomc package \citep{Lewis:2002ah} with the lensing module presented in \citet{joudaki16}. 

In our MCMC runs, we always vary the `vanilla' parameters $\left\{{\Omega_{\mathrm c}h^2, \Omega_{\mathrm b}h^2, \theta_{\rm MC}, n_{\mathrm s}, \ln{(10^{10} A_{\mathrm s})}}\right\}$, corresponding to the cold dark matter density, baryon density, approximation to the angular size of the sound horizon, scalar spectral index, and amplitude of the scalar spectrum, respectively, along with the optical depth to reionization, $\tau$, when including CMB measurements. The parameters $A_{\mathrm s}$ and $n_{\mathrm s}$ are defined at the pivot wavenumber $k_{\rm pivot}$. Moreover, we always vary the baryon feedback and intrinsic alignment amplitudes, $B$ and $A_{\rm IA}$ respectively, while the parameter governing the redshift dependence of the intrinsic alignment signal $\eta_{\rm IA}$ is varied in our `extended systematics' scenario.
Our treatment of the photometric redshift uncertainties does not involve any additional degrees of freedom.

We fiducially assume a flat universe and no running of the spectral index. Our fiducial cosmological model includes three massless neutrinos (adequate at the level of our constraints, negligible difference compared to assuming the 0.06~eV minimal mass of the normal hierarchy), such that the effective number of neutrinos $N_{\rm eff} = 3.046$. We determine the primordial helium abundance as a function of $N_{\rm eff}$ and $\Omega_{\mathrm b} h^2$ in a manner consistent with Big Bang Nucleosynthesis (BBN; see e.g.~equation~1 in \citealt{sj13}). The Hubble constant, $H_0$ (expressed as $h$ in its dimensionless form), and rms of the present linear matter density field on $8~h^{-1} {\rm Mpc}$ scales, $\sigma_8$, can be derived from the vanilla parameters. The uniform priors on the vanilla and lensing systematic parameters are listed in Table~\ref{table:priors}, which also contains the priors on the extended cosmology parameters discussed in Sections~\ref{neum} to \ref{runningsec}. 

As part of our MCMC computations, we use the \citet{Gelman92} $R$ statistic to determine the convergence of our chains, where $R$ is defined as the variance of chain means divided by the mean of chain variances. 
We enforce the conservative limit $(R - 1) < 2 \times 10^{-2}$, and stop the MCMC runs after further explorations of the distribution tails.

\section{Results}
\label{Results}

We now investigate the KiDS-450 extended systematics and cosmology constraints. In addition to a more conservative treatment of the intrinsic galaxy alignments, baryon feedback, the cosmological extensions considered are the sum of active neutrino masses, spatial curvature, evolving dark energy (both in the form of a constant equation of state and in the form of a time-dependent parameterization), evolving dark energy with curvature, modified gravity, and running of the scalar spectral index. 

The relative impact of these extensions on the lensing observables are shown in Figure~\ref{figcorr}.
We consider the relative preference of these extended models as compared to the standard model in Table~\ref{table:chidic}, and the impact of the extensions on the relative concordance between KiDS and Planck in Table~\ref{table:conc}. We only determine the joint KiDS+Planck parameter constraints in the event the two datasets are not in tension. Our criterion for this is $\log \mathcal{I} > 0$.

\begin{figure}
\hspace{-1.4em}
\resizebox{8.7cm}{!}{\includegraphics{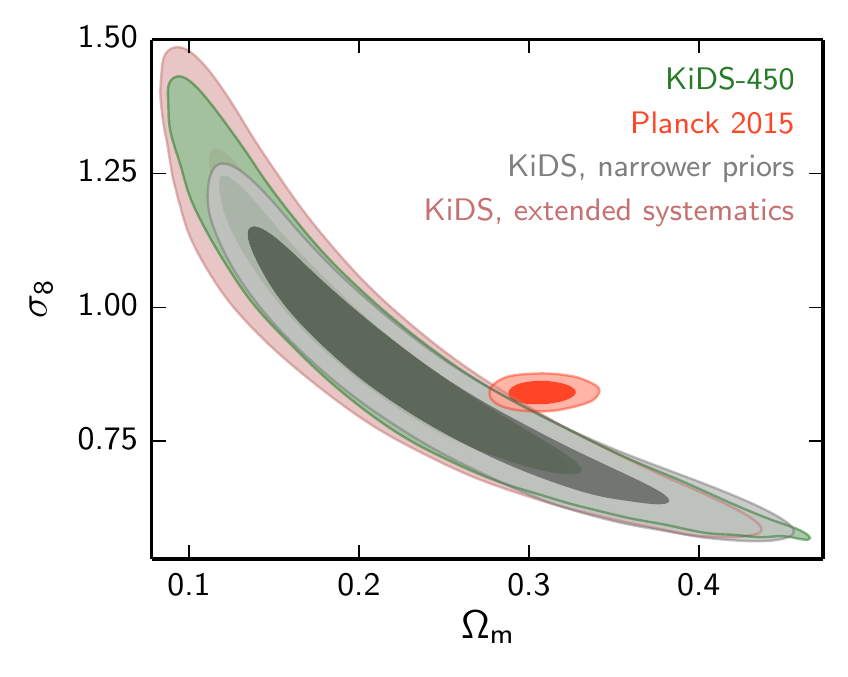}}
\vspace{-1.5em}
\caption{Marginalized posterior contours in the $\sigma_8 - \Omega_{\mathrm m}$ plane (inner 68\%~CL, outer 95\%~CL). We show our fiducial KiDS constraints in green, KiDS with narrower priors on the Hubble constant and baryon density in grey (as in \citealt{Hildebrandt16}), KiDS with extended treatment of the astrophysical systematics in pink, and Planck in red. 
}
\label{figh0prior}
\end{figure}

\subsection{$\Lambda$CDM (extended lensing systematics)}
\label{lcdmsec}

In \citet{Hildebrandt16}, we employed informative priors on the Hubble constant and baryon density ($\pm5\sigma$ of the constraints in \citealt{riess16} and \citealt{cyburt16}, respectively), but here we consider less informative priors on these parameters, in accordance with Table~\ref{table:priors}, as we move away from the fiducial $\Lambda$CDM model.

In Figure~\ref{figh0prior}, we show the cosmological constraints from KiDS in the $\sigma_8 - \Omega_{\mathrm m}$ plane, both using the same parameter priors as in \citet{Hildebrandt16}, and then widening the priors on the Hubble constant and baryon density in accordance with Table~\ref{table:priors}. As previously noted in \citet{joudaki16} and \citet{Hildebrandt16}, wider priors mainly extend the lensing contours along the degeneracy direction, and do not remove the tension with Planck. 
Thus, for both choices of priors, the tension between KiDS weak lensing and Planck CMB temperature (TT+lowP) measurements is approximately $2\sigma$, when quantified via the $S_8 = \sigma_8 (\Omega_{\mathrm m}/0.3)^{0.5}$ parameter combination that lensing measures well. Accounting for the full parameter space, we find $\log \mathcal{I} = -0.63$ (defined in Section~\ref{modsec}, and shown in Table~\ref{table:conc}), which corresponds to `substantial discordance' between the KiDS and Planck datasets. 
This is similar to the value $\log \mathcal{I} = -0.79$ found in \citet{Hildebrandt16}, despite the different priors on the Hubble constant and baryon density.

\begin{figure*}
\includegraphics[width=0.74\hsize]{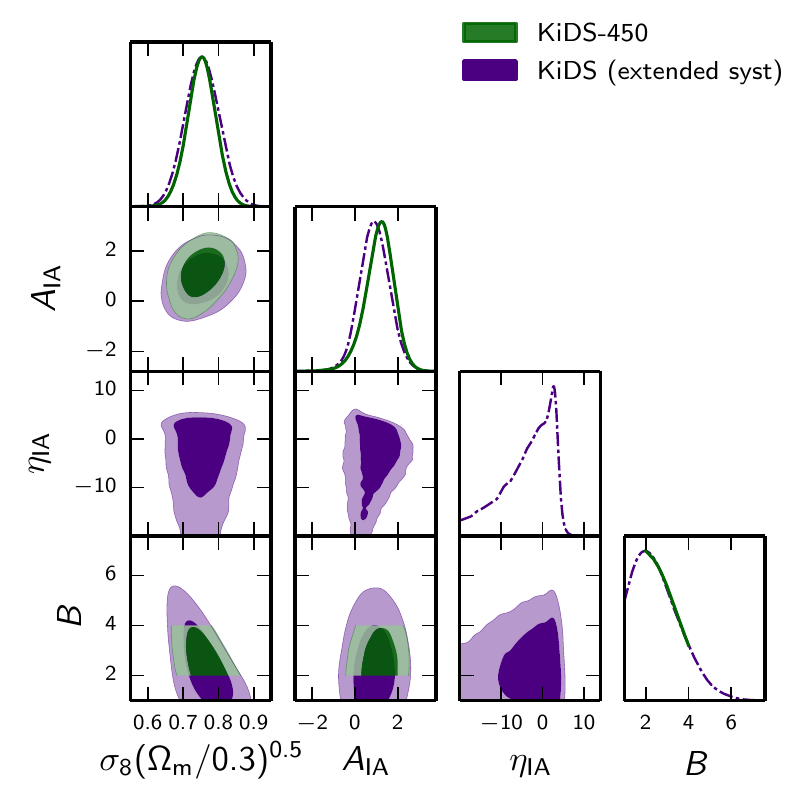}
\vspace{-2.0em}
\caption{\label{figsub} Marginalized posterior distributions of the lensing systematics parameters and their correlation. The vanilla parameters are simultaneously included in the analysis. We show KiDS with the fiducial treatment of systematic uncertainties in green (solid), and KiDS with the extended treatment of the lensing systematics in purple (dot-dashed). Parameter definitions and priors are listed in Table~\ref{table:priors}.}
\end{figure*}

We also examine the robustness of our fiducial treatment of the systematic uncertainties in KiDS, by allowing for a possible redshift dependence of the intrinsic alignment signal (via $\eta_{\rm IA}$), and simultaneously widening the priors on the intrinsic alignment amplitude, $A_{\rm IA}$, and baryon feedback amplitude $B$ entering \hmcode. Extending the prior on $B$ allows us to consider a greater range of feedback models. As some of the feedback models considered in the latest OWL simulations (cosmo-OWLS; \citealt{lebrun14}) are more extreme in the violence they inflict on the matter power spectrum than those in the original OWLS models \citep{Schaye10,Daalen11}, extending to low values of $B$ is an attempt to encompass this greater range of behaviours. 

We follow the strategy adopted in \citet{Hildebrandt16} to account for uncertainties in the multiplicative shear calibration correction and in the source redshift distributions. The analysis of \citet{fc16} showed that the shear calibration for KiDS is accurate at the level of $\lesssim1\%$, an error that is propagated~by~modifying the data covariance matrix (see equation 12 in \citealt{Hildebrandt16}). We used a range of different methods in \citet{Hildebrandt16} to validate the `DIR' calibrated redshift distributions that we adopt, and use bootstrap realizations of the set of tomographic redshift distributions to propagate our uncertainty on this redshift measurement through to cosmological parameter constraints (further see Section 6.3 of \citealt{Hildebrandt16}). We note that the accuracy of this redshift calibration method will continue to improve with the acquisition of additional spectroscopic redshifts to reduce the sample variance, which we estimate to be subdominant for KiDS-450 (see Appendix C3.1 in \citealt{Hildebrandt16}).

We are confident that this approach correctly propagates the known measured uncertainty in the multiplicative shear calibration correction and source redshift distributions but recognize that there could always be sources of systematic uncertainty that are currently unknown to the weak lensing community. Appendix A of \citet{Hildebrandt16} presents a Fisher matrix analysis that calculates how increasing the uncertainty on the shear calibration or redshift distribution results in an increase in the error on $S_8$. In our Appendix A, we verify the results of the Fisher matrix analysis by repeating our MCMC analysis allowing for an arbitrarily chosen Gaussian uncertainty of $\pm 10\%$ on the amplitudes of each of the tomographic shear correlation functions. The addition of these four new nuisance parameters could represent an unknown additional uncertainty in one or both of the shear and redshift calibration corrections. We find that the addition of these arbitrary nuisance parameters increases the error on $S_8$ by 15\% in agreement with the Fisher matrix analysis of \citet{Hildebrandt16}.

As shown in Figure~\ref{figcorr} (also see \citealt{semboloni11,semboloni13,joudaki16}), 
the baryon feedback suppresses the shear correlation functions on small angular scales across all tomographic bins, with a greater amount for a given angular scale in $\xi_{-}^{ij}(\theta)$ than in $\xi_{+}^{ij}(\theta)$. The suppression is larger in $\xi_{-}^{ij}(\theta)$ than $\xi_{+}^{ij}(\theta)$ because the former is more sensitive to nonlinear scales in the matter power spectrum for a given angular scale. By contrast, the intrinsic alignments mainly suppress the cross-tomographic bins, fairly uniformly across angular scale, and by approximately the same amount in $\xi_{+}^{ij}(\theta)$ as in $\xi_{-}^{ij}(\theta)$. The impact of a negative $\eta_{\rm IA}$ is to diminish the intrinsic alignment signal with increasing redshift, while a positive value boosts the intrinsic alignments with increasing redshift.

In Figure~\ref{figh0prior}, we find that the combined effect of the extensions in the lensing systematics modeling on the KiDS contour in the $\sigma_8 - \Omega_{\mathrm m}$ plane is small, as the contour mildly expands in a region of high $\sigma_8 $ and low $\Omega_{\mathrm m}$ where Planck is not located. The discordance between KiDS and Planck remains approximately the same, at the level of $1.8\sigma$ in $S_8$, and with  $\log \mathcal{I} = -0.70$. The slight decrease in the $S_8$ tension is not due to a noticeable shift in the KiDS estimate, but instead due to a 25\% increase in the uncertainty of the marginalized $S_8$ constraint (which picks up contributions from the widened contour in the full $\sigma_8 - \Omega_{\mathrm m}$ plane, even away from the Planck contour).

In Figure~\ref{figsub}, we show a triangle plot of the constraints in the subspace of the extended systematics parameters ($A_{\rm IA}, \eta_{\rm IA}, B$) along with $S_8$. We constrain the baryon feedback amplitude $B < 4.6$ (or $\log B < 0.66$) at 95\% confidence level (CL), with a peak around $B = 2$, which most closely corresponds to the `AGN' case in \citet{Mead15}.
We constrain the intrinsic alignment redshift dependence to be consistent with zero, where $-16 < \eta_{\rm IA} < 4.7$ (95\% CL). Although the posterior peaks for $\eta_{\rm IA} \gtrsim 0$, it has a sharp cutoff in the positive domain (as it boosts the IA signal and decreases the total lensing signal) and a long tail in the negative domain (as it diminishes the IA signal and does not contribute to the total lensing signal).

\begin{figure}
\hspace{-0.9em}
\resizebox{8.7cm}{!}{\includegraphics{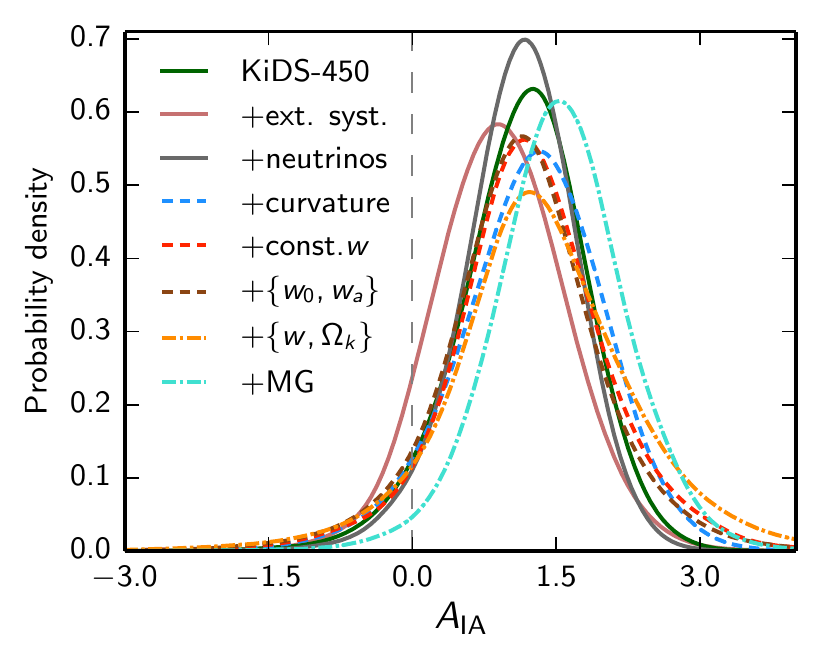}}
\vspace{-2.6em}
\caption{Marginalized posterior distributions for the intrinsic alignment amplitude considering different extended models. 
}
\label{figampia}
\end{figure}

Despite the redshift dependent degree of freedom, we continue to find an almost $2\sigma$ preference for a nonzero intrinsic alignment amplitude, where $-0.45 < A_{\rm IA} < 2.3$, which is similar to our constraint of $-0.24 < A_{\rm IA} < 2.5$ when considering the fiducial treatment of the systematic uncertainties. Both of these constraints are included in Figure~\ref{figampia}, which shows that the IA amplitude posteriors are remarkably consistent regardless of the systematic uncertainties and underlying cosmological model (discussed in forthcoming sections). Given the different imprints on the lensing observables, we~find no significant correlation between the intrinsic alignment~and baryon feedback parameters in Figure~\ref{figsub}. However, we do find a weak correlation between $S_8$ and the feedback amplitude.

In Table~\ref{table:chidic}, we show that although the extended systematics model improves the fit to the KiDS measurements by $\Delta\chi^2 = -2.1$ as compared to the fiducial model, it is marginally disfavored by $\Delta{\rm DIC} = 2.4$. Thus, in addition to not noticeably improving the discordance with Planck, extending the treatment of the systematic uncertainties in KiDS is marginally disfavored as compared to the fiducial treatment of the systematic uncertainties. We therefore also consider a `large-scale' cut, where we follow the approach in \citet{planckmg15} by removing all angular bins in the KiDS measurements except for the two bins centered at $\theta = \{24.9, 50.7\}$ arcmins in $\xi_{+}^{ij}(\theta)$, and the one bin centered at $\theta = 210$ arcmins in  $\xi_{-}^{ij}(\theta)$. The downsized data vector consists of 30 elements (from the fiducial 130 elements), and the angular scales that are kept are effectively insensitive to any nonlinear physics in the matter power spectrum, as for example seen for the case of baryons in Figure~\ref{figcorr}. However, the substantial discordance with Planck persists despite the removal of small scales in the lensing measurements, where $\log \mathcal{I} = -0.62$, and $T(S_8) = 1.9\sigma$ (as $S_8 = 0.55^{+0.32}_{-0.29}$ at 95\%~CL decreases away from Planck but has larger uncertainty).

In addition to changes in the treatment of the weak lensing systematic uncertainties and removal of small angular scales in the KiDS measurements, the tension with Planck is also robust to
changes in the choice of the CMB measurements. Including small-scale polarization information (Planck TT, TE, EE+lowP) increases the tension by another $0.2\sigma$, while including CMB lensing measurements (Planck TT+lowP+lensing) decreases the tension by roughly the same amount. 
Given our inability to resolve the discordance between KiDS and Planck in the context of the standard $\Lambda$CDM model, we therefore proceed by turning our attention to extensions to the underlying cosmological
model (with fiducial treatment of the systematic uncertainties), and examine to what extent these cosmological models are favored by the data while simultaneously alleviating the discordance between the two datasets.

\subsection{Neutrino mass}
\label{neum}

As we explore extensions to the standard model of cosmology, we begin by allowing for the sum of neutrino masses to vary as a free parameter in our MCMC analysis. 
Since massive neutrinos suppress the clustering of matter below the neutrino free-streaming scale, we need to adequately account for this in our estimation of the matter power spectrum over a range of redshifts and scales. 

To this end, we use the updated \citet{Mead16} version of \hmcode which can account for the impact of massive neutrinos on the nonlinear matter power spectrum in tandem with other physical effects, such as baryonic feedback. \hmcode is a tweaked version of the halo model, and as such the non-linear matter power spectrum it predicts responds to new physical effects in a reasonable way, even without additional calibration. 
To improve an already good match to the massive neutrino simulations of \citet[which assume a degenerate hierarchy with sum of neutrino masses between $0.15$~eV to $0.60$~eV]{massara14}, two physically motivated free parameters were introduced in \citet{Mead16} that were then calibrated to these simulations.
The updated \hmcode prescription matches the massive neutrino simulations at the few percent level (in the tested range $z \leq 1$ and $k \leq 10~h~{\rm{Mpc}}^{-1}$), which is a minor improvement compared to the fitting formula of \citet{bird12}, but with the additional benefit of simultaneously accounting for the impact of baryons.

\begin{figure*}
\begin{center}
\resizebox{8.8cm}{!}{\includegraphics{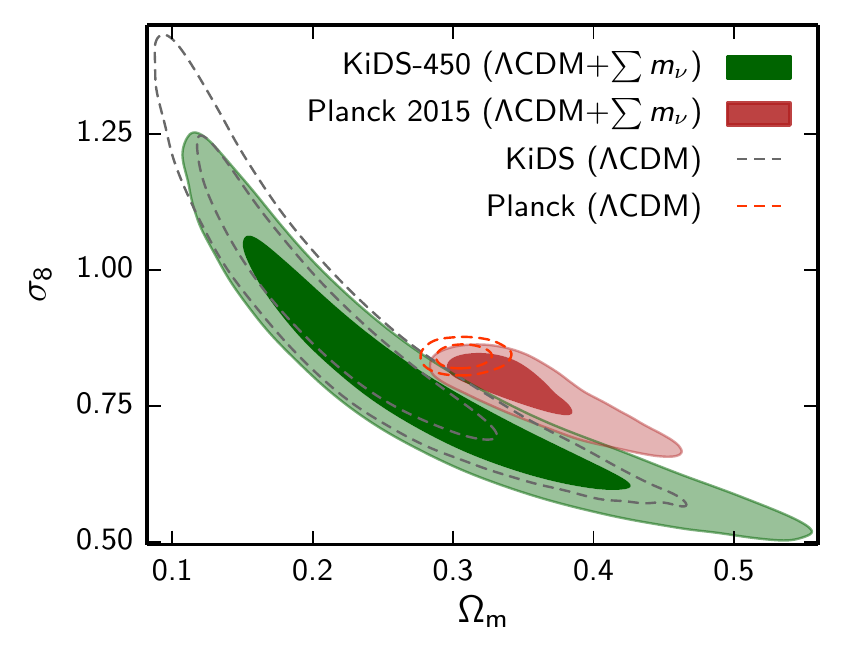}}
\resizebox{8.7cm}{!}{\includegraphics{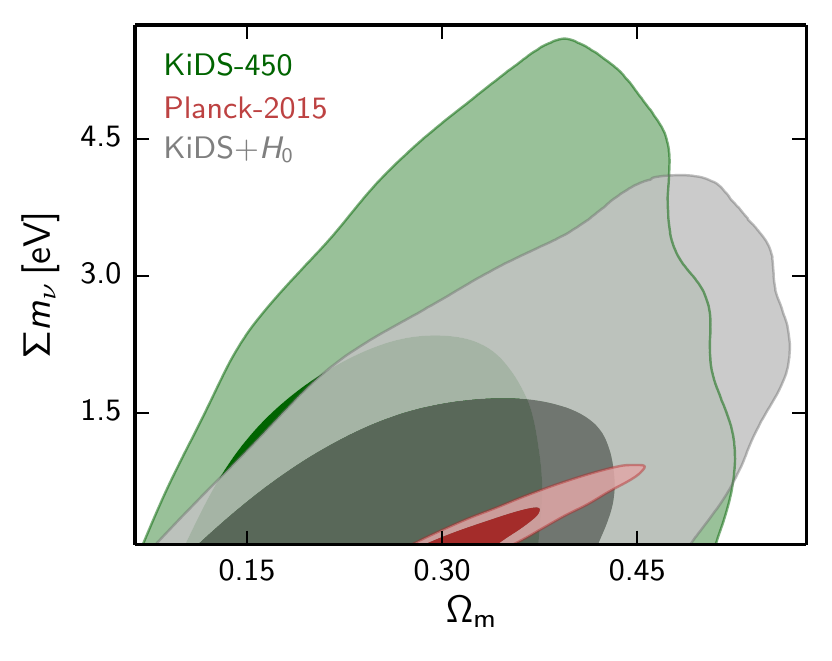}}
\end{center}
\vspace{-2.0em}
\caption{Left: Marginalized posterior contours in the $\sigma_8 - \Omega_{\mathrm m}$ plane (inner 68\%~CL, outer 95\%~CL) in a universe with massive neutrinos for KiDS in green and Planck in red. For comparison, dashed contours assume fiducial $\Lambda$CDM. Right: Marginalized posterior contours in the $\sum m_{\nu} - \Omega_{\mathrm m}$ plane for KiDS in green, KiDS with informative $H_0$ prior in grey (from \citealt{riess16}), 
and Planck in red.
}
\label{fignumass}
\end{figure*}

In Figure~\ref{figcorr}, we show the impact of three neutrinos with degenerate masses adding up to 1 eV on the shear correlation functions when using \hmcode for the modeling of the nonlinear matter power spectrum. As expected, the neutrino masses suppress the shear correlation functions on small angular scales, at roughly the same level across tomographic bins, and at a greater level in $\xi_{-}^{ij}(\theta)$ as compared to $\xi_{+}^{ij}(\theta)$, as the former is more sensitive to nonlinear scales in the matter power spectrum. In massive neutrino simulations, one finds that the matter power spectrum with massive neutrinos receives a boost beyond $k \approx 1~h/{\rm Mpc}$ (e.g. see Figure~3 in \citealt{Mead16}). We observe this `spoon-like' feature in the $\xi_{-}^{ij}(\theta)$ ratio within the angular scales probed by KiDS, and more prominently in the small-scale region that has been masked out. This indicates that probing these small scales (and beyond) could better help to disentangle the imprints of massive neutrinos from that of baryons (also see e.g.~\citealt{maccrann16}). 

In Figure~\ref{fignumass}, we show constraints in the $\sigma_8 - \Omega_{\mathrm m}$ and $\sum m_{\nu} - \Omega_{\mathrm m}$ planes. We continue to assume a degenerate neutrino mass hierarchy (adequate at the level of our constraints, also see e.g.~\citealt{hc12}), 
with the sum of neutrino masses as a free parameter in addition to the standard five $\Lambda$CDM parameters and two weak lensing systematics parameters ($A_{\rm IA}$ and $B$, all listed in Table~\ref{table:priors}). Allowing for the neutrinos to have mass pushes both the KiDS and Planck contours towards larger values of $\Omega_{\mathrm m}$ and smaller values of $\sigma_8$, but only along the degeneracy direction. Thus, although the KiDS and Planck contours are in greater contact, the tension in $S_8$ remains high at $2.4\sigma$. On the other hand, accounting for the full parameter space, we find $\log \mathcal{I} = -0.011$, which indicates there~is~neither discordance or concordance between the two datasets.

In the right hand panel of Figure~\ref{fignumass}, we find that the KiDS dataset is not sufficiently powerful to provide a strong bound on the sum of neutrino masses, with $\sum m_{\nu} < 4.0$~eV at 95\% CL 
(consistent with the power spectrum analysis in K\"ohlinger et al., in preparation). 
By imposing a uniform $\pm5\sigma$ prior on the Hubble constant from \citet{riess16}, the KiDS constraint improves to $\sum m_{\nu} < 3.0$~eV (95\% CL).
If one were to combine KiDS with Planck (given $\log \mathcal{I} \approx 0$), the addition of KiDS would only improve the Planck constraint on the sum of neutrino masses by 20\% (such that $\sum m_{\nu} < 0.58$~eV at 95\%~CL). As shown in Figure~\ref{figampia}, the constraint on the intrinsic alignment amplitude in this extended cosmology is only marginally affected by the inclusion of neutrino mass as a free parameter in our analysis, where $-0.12 < A_{\rm IA} < 2.3$ (95\% CL). If one were to combine KiDS with Planck (again as $\log \mathcal{I} \approx 0$), the constraint would improve to $0.43 < A_{\rm IA} < 2.0$ (95\% CL).

\begin{figure}
\hspace{-0.73em}
\resizebox{8.66cm}{!}{\includegraphics{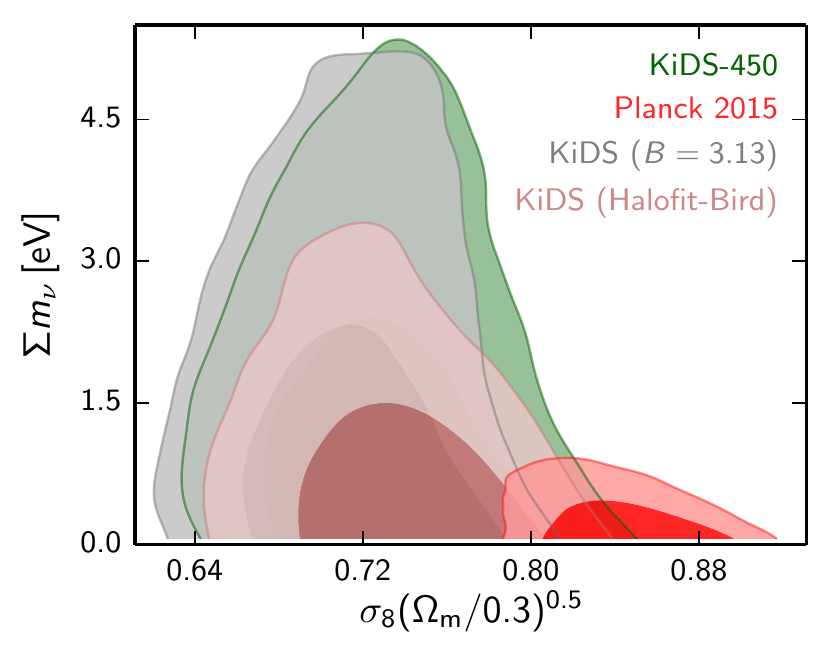}}
\vspace{-2.6em}
\caption{Marginalized posterior contours in the $\sum m_{\nu} - S_8$ plane (inner 68\%~CL, outer 95\%~CL). We show the results for KiDS in green with the fiducial treatment of baryons in \hmcode. We fix the feedback amplitude $B$ in \hmcode to its DM-only value in grey, we use \halofit instead of \hmcode in pink, and we consider Planck in red. 
}
\label{fignunu}
\end{figure}

Despite alleviating the discordance with Planck, the neutrino mass degree of freedom is not required by the data, as the difference in DIC relative to fiducial $\Lambda$CDM is 2.7 for KiDS, 3.4 for Planck, and 3.3 for KiDS+Planck. Moreover, the KiDS constraints on the sum of neutrino masses are not competitive with that of other data combinations; for instance, Planck with baryon acoustic oscillation (BAO) measurements from the 6dF Galaxy Survey \citep{beutler11}, SDSS Main Galaxy Sample \citep{ross15}, and BOSS LOWZ/CMASS samples \citep{anderson14} constrain $\sum m_{\nu} < 0.21$~eV at 95\% CL \citep{planck15}.

\begin{figure*}
\vspace{-1.5em}
\includegraphics[width=0.90\hsize]{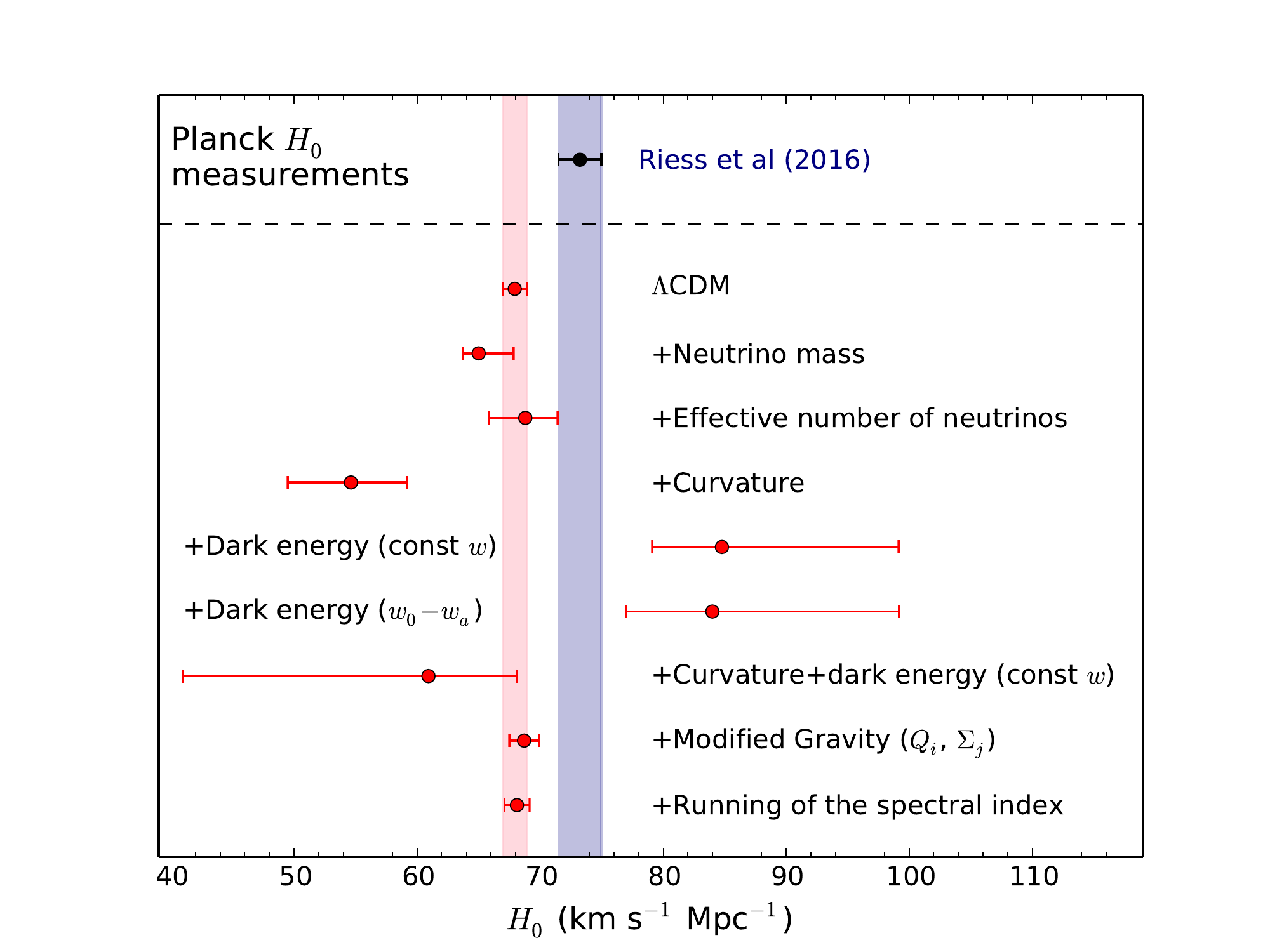}
\vspace{-0.7em}
\caption{Hubble constant constraints at 68\% CL in our fiducial and extended cosmologies, for Planck in red \citep{planck15} as compared to the direct measurement of \citet{riess16} in purple. We do not show the corresponding constraints for KiDS, as it is unable to measure the Hubble constant. Our $\Lambda$CDM constraint on the Hubble constant ($h = 0.679 \pm 0.010$) differs marginally from that in \citet[$h = 0.673 \pm 0.010$]{planck15} due to different priors, in particular our fiducial model fixes the neutrinos to be massless.
}
\label{fighubble}
\end{figure*}

In Figure~\ref{fignunu}, we show our neutrino mass constraints in the plane with $S_8$. We consider using \hmcode with the fiducial treatment of the baryon feedback amplitude as a free parameter (i.e.~corresponding to the same KiDS results in Figure~\ref{fignumass}), and we consider using \hmcode with the feedback amplitude fixed to $B = 3.13$ (along with fixing the bloating parameter to $\eta_\hmcode = 0.603$, in lieu of being determined by $B$), corresponding to a `DM-only' scenario. While the neutrino mass constraints are not significantly affected by these two different \hmcode scenarios, the KiDS constraint on $S_8$ is pushed further away from Planck when fixing the feedback amplitude to the DM-only value.

We compare the KiDS constraints in the  $\sum m_{\nu} - S_8$ plane to the case where the \halofit prescription (\citealt{Takahashi12, bird12}) is used to model the nonlinear matter power spectrum. 
Although \halofit, which is unable to account for the effect of baryonic physics in the nonlinear matter power spectrum, agrees well with \hmcode with DM-only settings, the KiDS neutrino mass bound with \halofit is stronger at $\sum m_{\nu} < 2.5$~eV (95\% CL). Moreover, the KiDS contour with \halofit is less in tension with Planck than when using \hmcode with DM-only settings, at a level of $2.5\sigma$ with \halofit as compared to $3.0\sigma$ with \hmcode. These differences in both neutrino mass constraint and discordance with Planck illustrate the importance of an accurate prescription for the modeling of the nonlinear matter power spectrum (also see \citealt{natarajan14}).

In Figure~\ref{fighubble}, we show how the Planck measurement of the Hubble constant changes as a function of the underlying cosmology. It is well known that the CMB temperature constraint on the Hubble constant is anti-correlated with the sum of neutrino masses (e.g.~\citealt{sj13, planck15}). The Planck measurement of the Hubble constant in a cosmology with $\sum m_{\nu}$ as a free parameter therefore shifts it further away from local measurements of $H_0$. The discordance between the Planck (TT+lowP) measurement of the Hubble constant ($h = 0.673 \pm 0.010$) and the local measurement in \citet[$h = 0.732 \pm 0.017$]{riess16} is $2.7\sigma$ in our fiducial $\Lambda$CDM cosmology with massless neutrinos. 
In a cosmology with $\sum m_{\nu}$ as a free parameter, this discordance increases with $0.599 < h < 0.689$ at 95\%~CL.

While the KiDS dataset is not particularly sensitive to the effective number of neutrinos $N_{\rm eff}$, we note that this additional degree of freedom does help to bring the Planck constraint on the Hubble constant in agreement with the direct measurement of \citet{riess16}. This is mainly achieved by widening the Planck error bars on the Hubble constant, such that $0.635 < h < 0.746$ (95\%~CL), with $N_{\rm eff} = 3.15 \pm 0.32$. 
However, Planck does not favor this additional degree of freedom, as $\Delta{\rm{DIC}} = 1.1$.

\begin{figure*}
\begin{center}
\resizebox{8.5cm}{!}{\includegraphics{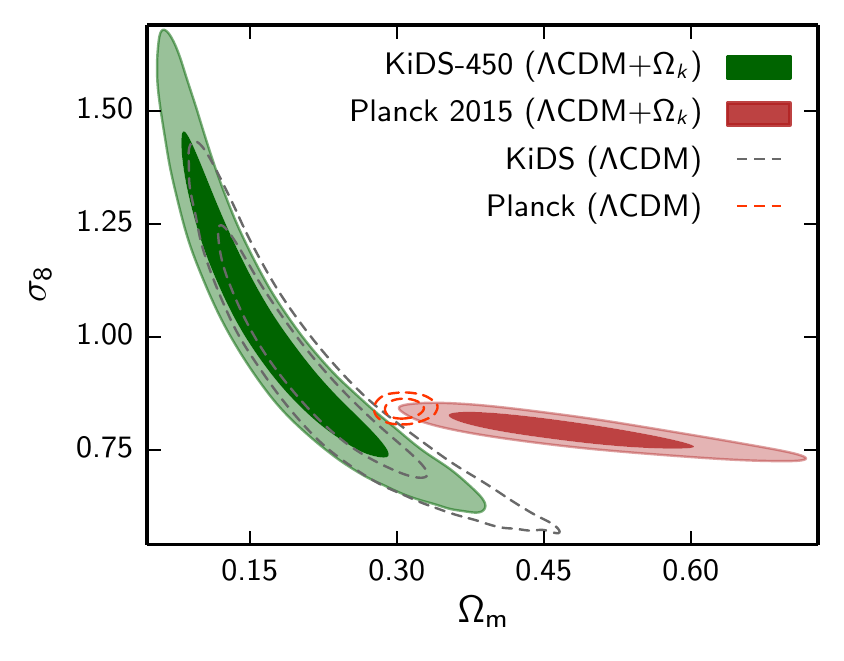}}
\resizebox{8.8cm}{!}{\includegraphics{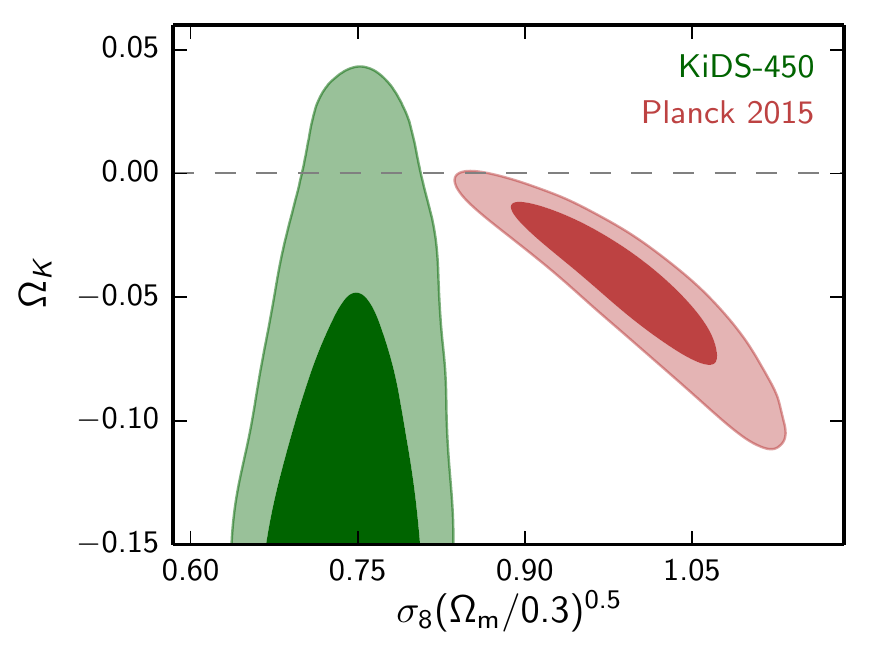}}
\end{center}
\vspace{-2.0em}
\caption{Left: Marginalized posterior contours in the $\sigma_8 - \Omega_{\mathrm m}$ plane (inner 68\%~CL, outer 95\%~CL) in a universe with nonzero curvature for KiDS in green and Planck in red. For comparison, dashed contours assume fiducial $\Lambda$CDM. Right: Marginalized posterior contours in the $\Omega_k - S_8$ plane for KiDS in green and Planck in red. The dashed horizontal line denotes flatness.
}
\label{figcurv}
\end{figure*}

\subsection{Curvature}
\label{curvsec}

We now move to constraining deviations from spatial flatness and examine the model selection and dataset concordance outcomes of this new degree of freedom for KiDS and Planck. 

In Figure~\ref{figcorr}, we show that a negative curvature (corresponding to a positive $\Omega_k$) decreases the shear signal, fairly uniformly across $\xi_{\pm}^{ij}(\theta)$ over the angular scales probed by KiDS, 
such that its signature can in principle be disentangled from that of lensing systematics such as baryons and intrinsic alignments. We note that when $\Omega_k$ is varied, $H_0$ is also varying to keep $\theta_{\mathrm{MC}}$ fixed (as the former is a derived parameter, while the latter is a primary parameter). If we vary the curvature by the same amount, and simultaneously vary $\theta_{\mathrm{MC}}$ such that $H_0$ is kept fixed instead, the decrease in the shear correlation functions 
reduces by almost an order of magnitude. Meanwhile, CMB temperature measurements of the curvature are highly correlated with the Hubble constant and matter density (due to their degeneracy in the angular diameter distance to the last scattering surface). The Planck constraint on the curvature mainly originates from the signatures of lensing in the CMB temperature power spectrum, the late-time integrated Sachs-Wolfe effect, and the lower boundary of the $H_0$ prior (e.g.~\citealt{wmap5,planck15}).

As a result, given that we exclude CMB lensing ($\phi\phi$), Planck is no longer able to constrain the matter density well when allowing $\Omega_k$ to vary, causing a nearly horizontal elongation of the Planck contour towards larger values of the matter density in the $\sigma_8 - \Omega_{\mathrm m}$ plane of Figure~\ref{figcurv} (and thereby larger $S_8$), while KiDS largely moves along the degeneracy direction towards smaller values of the matter density (with a minor offset that decreases $S_8$). 
The overall effect of these changes is to increase the tension between KiDS and Planck to $3.5\sigma$ in $S_8$ (where the main cause of the increased tension is the new Planck constraint,
which has shifted by a factor of six of the original uncertainty in $S_8$). 
Although Planck constrains $S_8$ more strongly than KiDS in a flat $\Lambda$CDM universe (by a factor of 1.7), the KiDS constraint on $S_8$ is a factor of 1.6 stronger than the constraint from Planck when $\Omega_k$ is allowed to vary.

Accounting for the full parameter space, $\log \mathcal{I} = -1.7$, which corresponds to `strong discordance' between the KiDS and Planck datasets.
In the $\Omega_k - S_8$ plane of Figure~\ref{figcurv}, the KiDS and Planck contours prefer $\Omega_k < 0$, both at approximately 95\% CL. Despite the deviation from flatness, the KiDS intrinsic alignment amplitude remains robustly determined as shown in Figure~\ref{figampia}, marginally widening to $-0.38 < A_{\rm IA} < 2.8$ (95\% CL). 
While Planck weakly-to-moderately favors nonzero curvature with $\Delta{\rm{DIC}} = -4.3$ (down from $\Delta \chi^2_{\rm eff} = -5.8$ due to the increased Bayesian complexity), the additional degree of freedom is not favored by KiDS, with $\Delta{\rm{DIC}} \simeq 0$.
Moreover, as shown in Figure~\ref{fighubble}, the Planck constraint on the Hubble constant ($0.46 < h < 0.65$ at 95\% CL) moves it further away from the \citet{riess16} result. 
Although the combination of weak lensing and CMB can significantly improve the constraint on the curvature (e.g.~\citealt{kilbinger13, planck15}), we do not provide joint KiDS+Planck constraints on $\Omega_k$ as the two datasets are discordant in this extended cosmology.

\subsection{Dark energy (constant $w$)}
\label{constwlab}

We now turn away from the assumption of a cosmological constant by considering evolving dark energy. We begin by allowing for a constant dark energy equation of state $w$ that can vary freely in our MCMC analyses.
While we have discussed \hmcode's ability to account for the impact of baryons and massive neutrinos in the nonlinear matter power spectrum, \hmcode's calibration to the Coyote N-body simulations also included models with $-0.7 < w < 1.3$ \citep{Mead15}. Our prior on $w$ extends beyond this range, but we expect our results to be only marginally biased, as the cosmological constraints are either too weak or tend to lie near $w = -1$. Moreover, in contrast to e.g. a fitting function, the physical grounding of \hmcode in the halo model allows one to probe fairly extreme values of $w$ and still trust the modeling, as changes to the underlying cosmology
diffuse through into the matter power spectrum prediction in a natural way (via the mass-concentration relation and evolution of the halo mass function). 

In Figure~\ref{figcorr}, we show the imprint of a constant dark energy equation of state on the shear correlation functions, while keeping all primary parameters fixed. An increase in the equation of state, such that $w > -1$, causes a scale-dependent suppression in the matter power spectrum relative to a cosmological constant (e.g.~\citealt{JK12, Mead16}). For a fixed Hubble constant, $w > -1$ also suppresses the lensing kernel relative to a cosmological constant (as it boosts $H(z)/H_0$), but this is not the case in Figure~\ref{figcorr} as $\theta_{\mathrm{MC}}$ is kept fixed in lieu of the Hubble constant which varies from one cosmology to another (since $\theta_{\mathrm{MC}}$ is a primary parameter while $H_0$ is treated as a derived parameter). 
Thus, when fixing our primary parameters, the lensing kernel increases for $w > -1$, partly canceling the suppression in the matter power spectrum.

\begin{figure*}
\begin{center}
\resizebox{8.8cm}{!}{\includegraphics{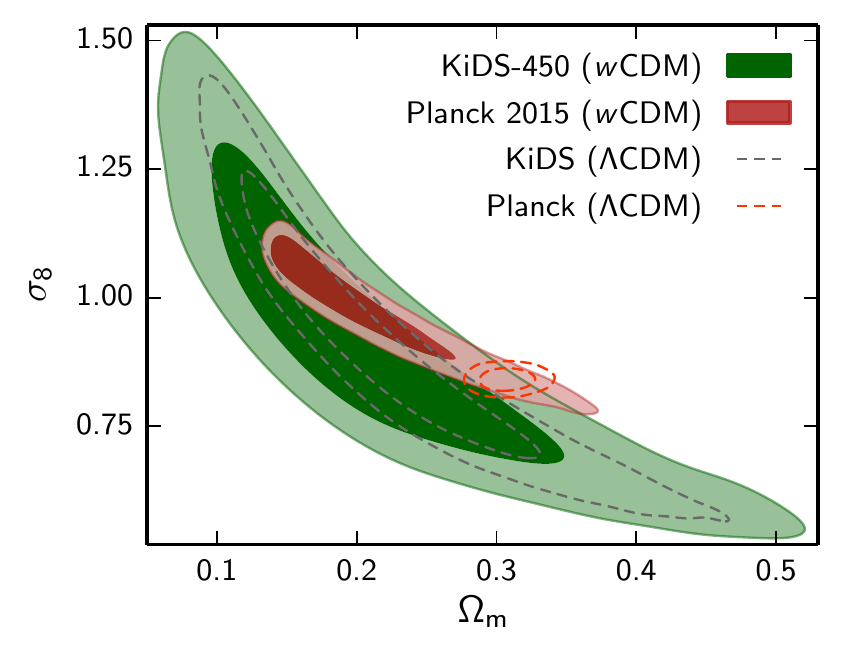}}
\resizebox{8.8cm}{!}{\includegraphics{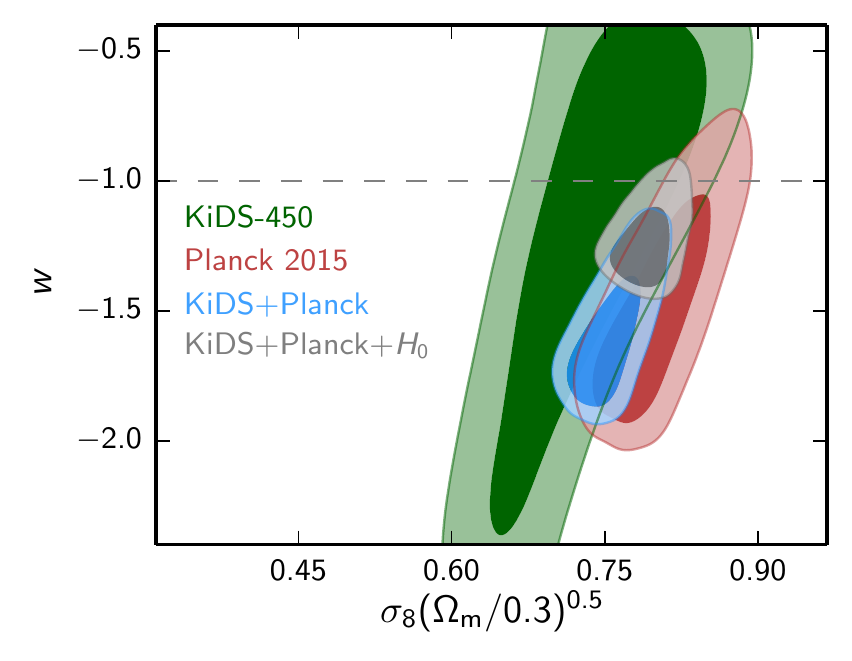}}
\end{center}
\vspace{-2.0em}
\caption{Left: Marginalized posterior contours in the $\sigma_8 - \Omega_{\mathrm m}$ plane (inner 68\%~CL, outer 95\%~CL) in a universe with a constant dark energy equation of state for KiDS in green and Planck in red. For comparison, dashed contours assume fiducial $\Lambda$CDM. Right: Marginalized posterior contours in the $w - S_8$ plane for KiDS in green, Planck in red, KiDS+Planck in blue, and KiDS+Planck with informative $H_0$ prior in grey (from \citealt{riess16}). The dashed horizontal line denotes the $\Lambda$CDM prediction.
}
\label{figconstw}
\end{figure*}

In Figure~\ref{figconstw}, we show the constraints in the $\sigma_8 - \Omega_{\mathrm m}$ and $w - S_8$ planes
when allowing for $w \neq -1$.
The KiDS and Planck contours now overlap in the $\sigma_8 - \Omega_{\mathrm m}$ plane, both due to a fairly uniform increase in the area of the KiDS contour perpendicular to the lensing degeneracy direction (noting that the lensing constraints parallel to the degeneracy direction are prior-dependent), and due to a shift in the Planck contour perpendicular to the lensing degeneracy direction. The realignment of the CMB contour along the lensing degeneracy direction was also found for CFHTLenS and WMAP7 in \citet{kilbinger13}, and the extension of the Planck contour along the $\Omega_{\mathrm m}$ axis is due to the same geometric degeneracy as in the case of a nonzero curvature.
As a result, the respective KiDS and Planck $S_8$ constraints agree at $1\sigma$ (despite seemingly being in tension in the $w-S_8$ plane). Accounting for the full parameter space, we find $\log \mathcal{I} = 0.99$, which effectively corresponds to `strong concordance' between the KiDS and Planck datasets.
In addition to removing the tension between these datasets, the Planck constraint on the Hubble constant is now also wider than in $\Lambda$CDM ($0.66 < h < 1.0$ at 95\%~CL, where the upper bound is hitting against the prior) and in agreement with the \citet{riess16} direct measurement of $H_0$.

In the $w-S_8$ plane, KiDS and Planck are both in agreement with a cosmological constant, while the combined analysis of KiDS+Planck seems to favor a $2.6\sigma$ 
deviation from $\Lambda$CDM (marginalized constraint of $-1.93 < w < -1.06$ at 99\%~CL). As noted in \citet{planck15}, deviations from a cosmological constant seem to be preferred by large values of the Hubble constant (that are arguably ruled out), and so we also consider a $\pm5\sigma$ uniform \citet{riess16} prior on $H_0$. While the KiDS+Planck+$H_0$ contour tightens and moves towards $w = -1$, we still find an approximately $2\sigma$ deviation from a cosmological constant (marginalized constraint of $-1.42 < w < -1.01$ at 95\%~CL). As in other extended cosmologies, the intrinsic alignment amplitude remains robustly determined when allowing $w$ to vary, with 95\% confidence levels at $-0.50 < A_{\rm IA} < 2.9$ for KiDS, $0.27 < A_{\rm IA} < 3.0$ for KiDS+Planck, and $0.38 < A_{\rm IA} < 2.4$ for KiDS+Planck+$H_0$.

\begin{figure*}
\begin{center}
\resizebox{8.63cm}{!}{\includegraphics{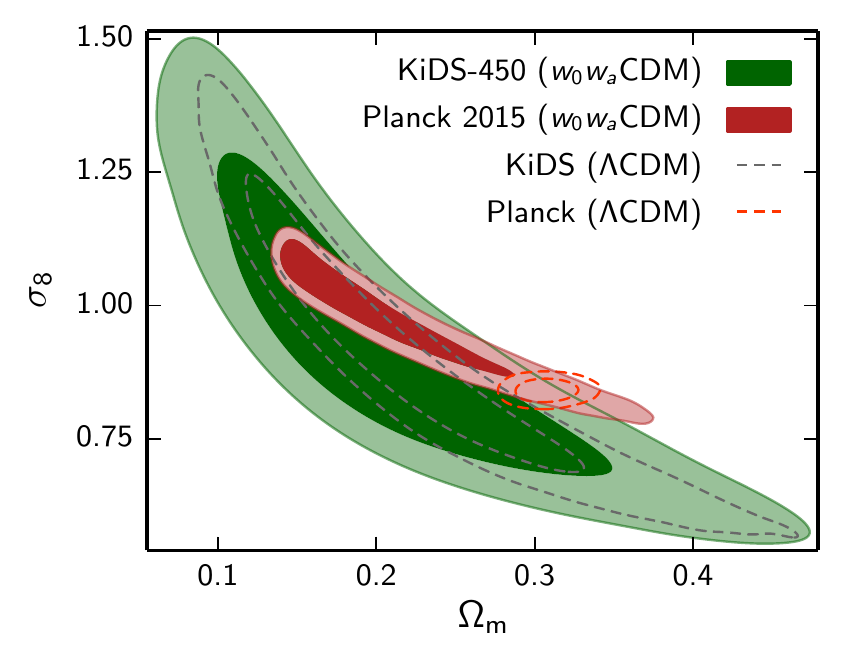}}
\resizebox{8.8cm}{!}{\includegraphics{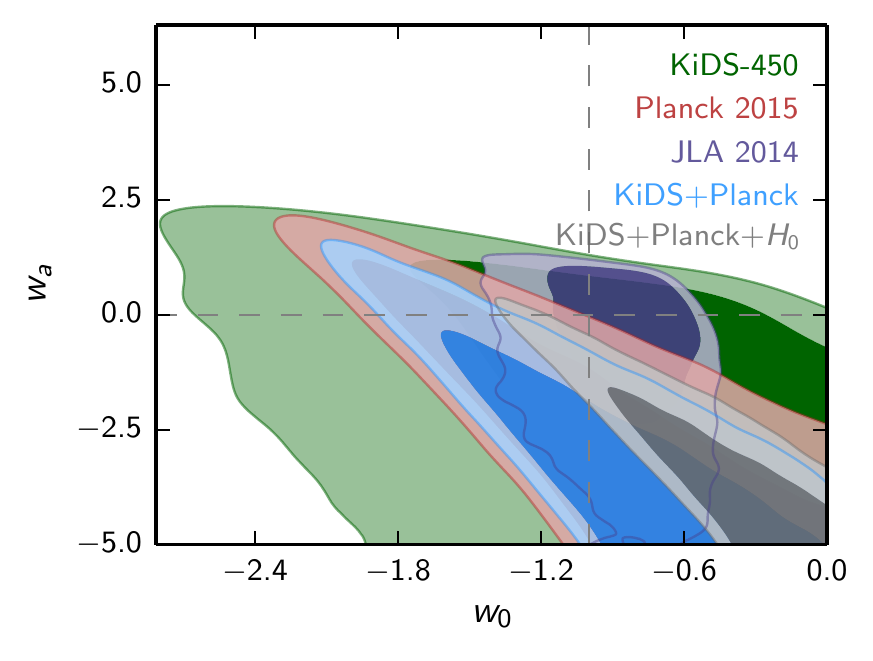}}
\end{center}
\vspace{-2.0em}
\caption{Left: Marginalized posterior contours in the $\sigma_8 - \Omega_{\mathrm m}$ plane (inner 68\%~CL, outer 95\%~CL) in a universe with a time-dependent dark energy equation of state for KiDS in green and Planck in red. For comparison, dashed contours assume fiducial $\Lambda$CDM. Right: Marginalized posterior contours in the $w_0 - w_a$ plane for KiDS in green, Planck in red, JLA SNe in purple, KiDS+Planck in blue, and KiDS+Planck with informative $H_0$ prior in grey (from \citealt{riess16}). The dashed lines denote the $\Lambda$CDM prediction.
}
\label{figw0wa}
\end{figure*}

We have shown that the introduction of a constant dark energy equation of state seems to remove the discordance between KiDS and Planck, and between local Hubble constant measurements and Planck, while moreover deviating from a cosmological constant when these measurements are combined. However, we also want to know to what extent the constant $w$ model is favored or disfavored by the data. We find that KiDS and Planck on their own show no preference for $w \neq -1$, with ${\Delta{\rm DIC}} = 2.3$ for KiDS and ${\Delta{\rm DIC}} = -0.20$ for Planck (respectively degraded from $\Delta \chi^2_{\rm eff} = 0.074$ and $\Delta \chi^2_{\rm eff} = -3.1$ due to the increased Bayesian complexity). However, the combination of KiDS+Planck seems to prefer the constant dark energy equation of state model with ${\Delta{\rm DIC}} = -5.4$ (with near identical Bayesian complexity to $\Lambda$CDM), while this preference reduces to ${\Delta{\rm DIC}} = -2.9$ when further considering KiDS+Planck+$H_0$ (marginally degraded from $\Delta \chi^2_{\rm eff} = -3.4$). Thus, from the point of model selection, we only find weak preference in favor of a constant dark energy equation of state model as compared to standard $\Lambda$CDM.

\subsection{Dark energy ($w_0$-$w_a$)}

Although a constant dark energy equation of state as a free parameter constitutes the simplest deviation from a $w = -1$ model, there is no strong theoretical motivation to keep the equation of state constant once one has moved away from the cosmological constant scenario. We therefore also consider a time-dependent parameterization to the equation of state, in the form of a first-order Taylor expansion with two free parameters: 
\begin{equation}
w(a) = w_0 + (1-a) w_a, 
\label{eqn:w0wa}
\end{equation}
where $a$ is the cosmic scale factor, $w_0$ is the dark energy equation of state at present, and 
$w_a = - {{\mathrm d}w}/{{\mathrm d}{a}}|_{a=1}$ (which can also be expressed as $w_a = -2 {{\mathrm d}w}/{{\mathrm d}\ln{a}}|_{a=1/2}$; \citealt{cp01}; \citealt{linder03}).

In Figure~\ref{figcorr}, we show the impact of a time dependence of the equation of state on the shear correlation functions. Since a negative $w_a$ makes the overall equation of state more negative with time, it has the opposite impact on the matter power spectrum and lensing kernel (and thereby shear correlation functions) to the case where $w > -1$ discussed in Section~\ref{constwlab}. Clearly the benefit of two degrees of freedom to describe the dark energy is that more complex behavior of the shear correlation functions is allowed than when only a constant equation of state is considered, enhancing the ability of the theoretical model to describe the data.
Meanwhile, the extra degree of freedom from nonzero $w_a$ further adds to the geometric degeneracy of the CMB measurements.

Along with the case where the dark energy equation of state is constant, \hmcode accurately accounts for the impact of $w_0 - w_a$ models on the nonlinear matter power spectrum, as demonstrated by the N-body simulations in \citet{Mead16}, covering $-1.0 < w_a < 0.75$ to $z \leq 1$ and $k \leq 10~h~{\rm{Mpc}}^{-1}$ (using a modified version of the \gadget code of \citealt{springel05}). 
\hmcode's excellent performance, which is similar to that of \halofit over the redshifts and scales considered, derives from the fact that the halo model is firmly grounded in physical reality. As a result, the non-linear power spectrum responds to cosmological extensions in a reasonable way via the linear growth, halo mass function, and halo mass-concentration relation, and has been shown to produce an excellent match to the non-linear response in simulations for a range of other dark energy models with a time-varying equation of state \citep{Mead16}.
For these reasons, we expect \hmcode to be adequate over our full prior range.

Using \hmcode to describe the nonlinear matter power spectrum, we constrain the two degrees of freedom $w_0$ and $w_a$ along with the vanilla and lensing systematics parameters (and CMB degrees of freedom when applicable). In Figure~\ref{figw0wa}, we show these constraints in the $\sigma_8 - \Omega_{\mathrm m}$ and $w_0 - w_a$ planes. 
Similar to the case where the equation of state is constant (Section~\ref{constwlab}), KiDS and Planck overlap in the $\sigma_8 - \Omega_{\mathrm m}$ plane, and are no longer in tension in the $S_8$ parameter ($1\sigma$ agreement). When accounting for the full parameter space, $\log \mathcal{I} = 0.82$, which corresponds to `substantial concordance' between the KiDS and Planck datasets. 
Moreover, as shown in Figure~\ref{fighubble}, the Planck constraint on the Hubble constant is wider than in $\Lambda$CDM ($0.65 < h < 1.0$ at 95\%~CL, where the upper bound is limited by the prior) and in agreement with the \citet{riess16} direct measurement of $H_0$. 
The KiDS constraint on the intrinsic alignment amplitude is marginally wider than in $\Lambda$CDM, with $-0.69 < A_{\rm IA} < 2.9$ (95\% CL), and this improves to $0.13 < A_{\rm IA} < 2.8$ (95\% CL) for KiDS+Planck, and $0.27 < A_{\rm IA} < 2.1$ (95\% CL) for KiDS+Planck+$H_0$.

\begin{figure}
\hspace{-0.53em}
\resizebox{8.7cm}{!}{\includegraphics{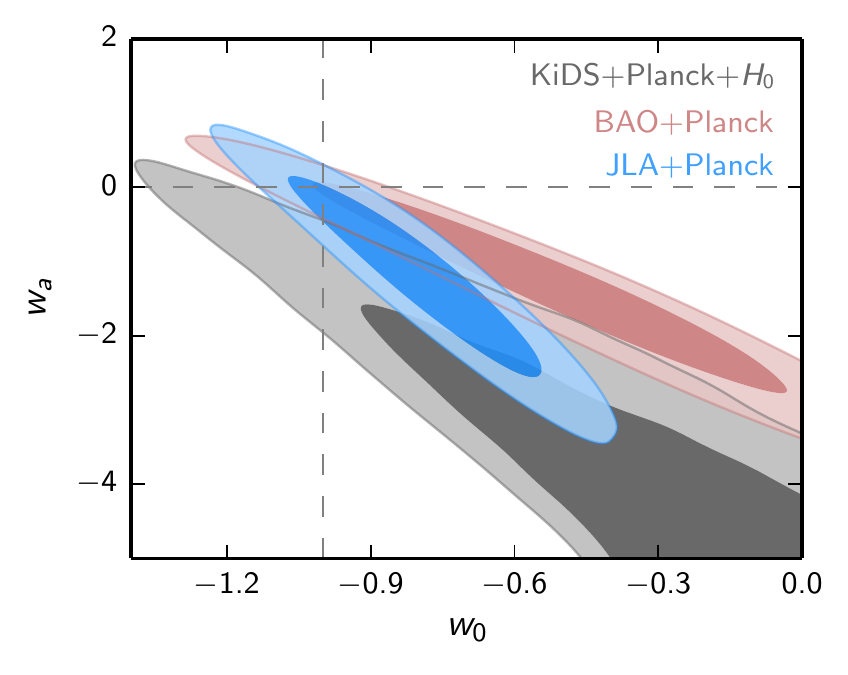}}
\vspace{-2.5em}
\caption{Marginalized posterior contours in the $w_0 - w_a$ plane (inner 68\%~CL, outer 95\%~CL) for Planck combined with weak lensing, BAO, and SN (JLA) measurements. We show the results for KiDS+Planck with a $\pm5\sigma$ uniform prior on the Hubble constant from \citet{riess16} in grey. We show BAO+Planck in pink, where the BAO measurements are from 6dFGS \citep{beutler11}, SDSS MGS \citep{ross15}, and BOSS LOWZ/CMASS samples \citep{anderson14}. We show JLA+Planck in blue, where the SN measurements are from \citet{Betoule13, Betoule14}.
}
\label{figw0wabaosne}
\end{figure}

When examining the constraints in the $w_0 - w_a$ plane, KiDS is in agreement with $\Lambda$CDM, while Planck shows an approximately $2\sigma$ deviation from a cosmological constant. Combining KiDS+Planck gives an even larger deviation from the cosmological constant scenario at $3.0\sigma$. Analogously to the constant $w$ case (and the discussion therein), imposing a Hubble constant prior pulls the KiDS+Planck+$H_0$ contour towards $\Lambda$CDM, but the prior also helps decrease the area of the error contour such that the statistical deviation from $\Lambda$CDM 
is still significant at approximately 3$\sigma$ (precisely, 2.7$\sigma$). 
This seeming preference of KiDS+Planck for evolving dark energy is consistent with the supernova distance measurements of the `Joint Light-curve Analysis' sample~(JLA, constructed from SDSS-II, SNLS, and low-redshift samples of SN data, \citealt{Betoule13,Betoule14}), and can be contrasted with the CFHTLenS+Planck scenario, where \citet{planck15} found that a Hubble constant prior is sufficient to bring the CFHTLenS+Planck results in agreement with $\Lambda$CDM. 

\begin{figure*}
\begin{center}
\resizebox{8.8cm}{!}{\includegraphics{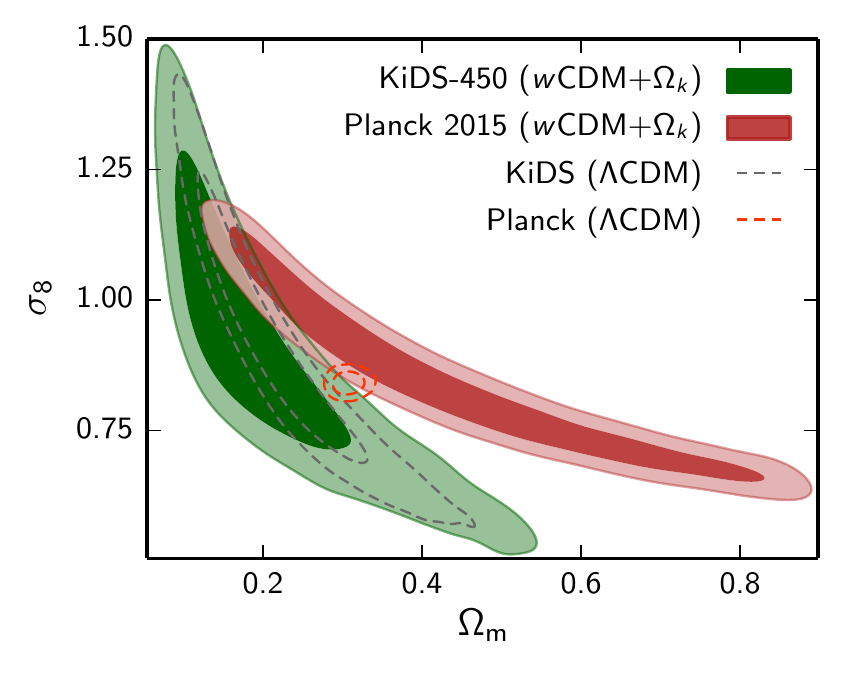}}
\resizebox{8.8cm}{!}{\includegraphics{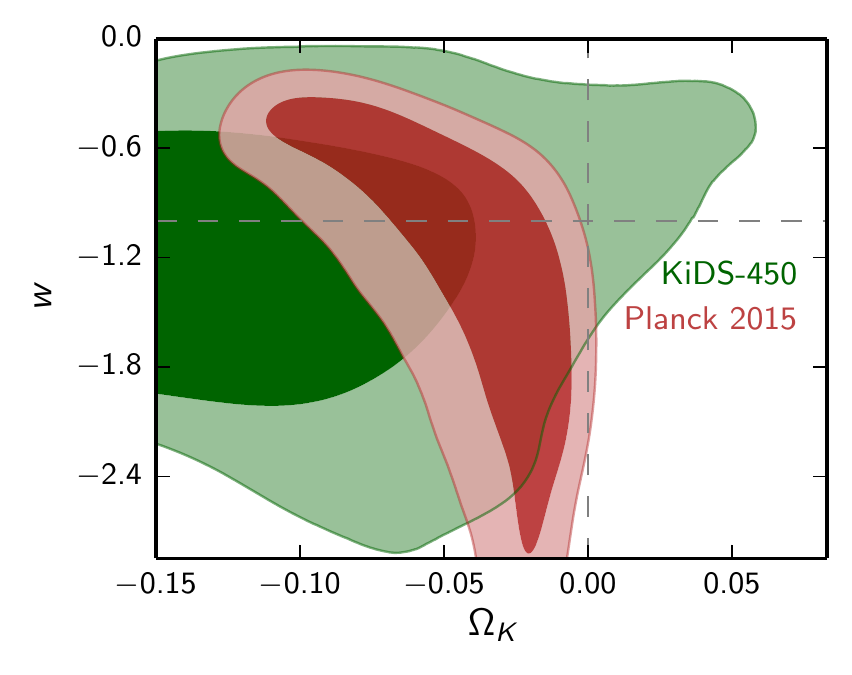}}
\end{center}
\vspace{-2.0em}
\caption{Left: Marginalized posterior contours in the $\sigma_8 - \Omega_{\mathrm m}$ plane (inner 68\%~CL, outer 95\%~CL) in a universe with both nonzero curvature and constant dark energy equation of state for KiDS in green and Planck in red. For comparison, dashed contours assume fiducial $\Lambda$CDM. Right: Marginalized posterior contours in the $w - \Omega_k$ plane for KiDS and Planck (green and red, respectively). The dashed horizontal line denotes the cosmological constant prediction, while the dashed vertical line denotes flatness.
}
\label{figwcurv}
\end{figure*}

Given the $3\sigma$ deviation from $\Lambda$CDM, in Figure~\ref{figw0wabaosne} we examine to what extent the KiDS+Planck+$H_0$ constraints in the $w_0 - w_a$ plane are consistent with the constraints from other probes combined with Planck. To this end, Planck is combined with SNe from JLA, and BAOs from the 6dF Galaxy Survey \citep{beutler11}, SDSS Main Galaxy Sample \citep{ross15}, and BOSS LOWZ/CMASS samples \citep{anderson14}. In the $w_0 - w_a$ plane, KiDS+Planck+$H_0$ is seemingly in tension with BAO+Planck, and in agreement with JLA+Planck (which also partly overlaps with BAO+Planck). While all three data combinations seem to be pulled towards $\{w_0 > -1, w_a < 0\}$, BAO+Planck and JLA+Planck are consistent with a cosmological constant at 95\% CL. In this extended cosmology, the constraint on the Hubble constant from JLA+Planck is $0.66 < h < 0.74$ (95\% CL), in agreement with the measurement from \citet{riess16}, while the constraint from BAO+Planck is $0.59 < h < 0.69$ (95\% CL), in tension with the measurement from \citet{riess16}. Thus, it seems difficult to reconcile all the measurements simultaneously when combined with Planck. Meanwhile, the constraints from KIDS+BAO and KIDS+JLA are weaker, in agreement both with KiDS+Planck+$H_0$ and with a cosmological constant.

The next step is to examine to what extent the two dark energy degrees of freedom 
are favored or disfavored by the KiDS and Planck datasets as compared to a cosmological constant from the point of model selection. Employing again the deviance information criterion, there is no preference away from $\Lambda$CDM for KiDS and Planck on their own (${\Delta{\rm DIC}} = 0.95$ for KiDS and ${\Delta{\rm DIC}} = -1.1$ for Planck, respectively degraded from $\Delta \chi^2_{\rm eff} = -0.35$ and $\Delta \chi^2_{\rm eff} = -3.2$ due to the increased Bayesian complexity). However, when KiDS and Planck are combined, there is moderate preference in favor of the $w_0-w_a$ model as compared to $\Lambda$CDM, with ${\Delta{\rm DIC}} = -6.4$ (marginally degraded from $\Delta \chi^2_{\rm eff} = -6.8$).
In contrast to the constant $w$ case in Section~\ref{constwlab}, this preference for evolving dark energy remains when further including the~\citet{riess16}~prior on the Hubble constant, such that ${\Delta{\rm DIC}} = -6.5$ for KiDS+Planck+$H_0$ (with similar Bayesian complexity to $\Lambda$CDM). 
Thus, from the point of model selection, there seems to be moderate preference in favor of the extended model when restricting the $H_0$ space in combining KiDS and Planck.

\subsection{Curvature + dark energy (constant $w$)}

In previous sections, we have considered unitary extensions to the standard cosmological model, in the form of neutrino mass, curvature, and dark energy. But the impact of these extensions on the cosmological observables are often correlated (e.g.~Figure~\ref{figcorr}), and we therefore also consider a simple combination of curvature and dark energy with a constant equation of state. In other words, we simultaneously vary the curvature density parameter $\Omega_k$ and dark energy equation of state $w$ in addition to the vanilla and lensing systematics parameters (along with the CMB degrees of freedom when applicable).

In Figure~\ref{figwcurv}, we show our constraints in the $\sigma_8 - \Omega_{\mathrm m}$ and $w - \Omega_k$ planes. In previous sections, we found that allowing for nonzero curvature increases the discordance between KiDS and Planck, while evolving dark energy increases the concordance between the datasets. 
In a cosmology with both $\Omega_k$ and $w$, the two parameters therefore partially cancel in their combined impact on the level of concordance between KiDS and Planck. 
In the $\sigma_8 - \Omega_{\mathrm m}$ plane, it is evident that Planck's ability to constrain the matter density is further degraded as compared to the unitary curvature and dark energy extensions to $\Lambda$CDM (due to the geometric degeneracy of the CMB), stretching over large parts of the parameter space where there is no overlap with KiDS. Although the area of the KiDS contour mainly expands away from Planck, the two contours partly overlap for small values of the matter density and large values of $\sigma_8$. Examining the tension in the marginalized $S_8$ constraints, $T(S_8) = 2.5\sigma$, while accounting for the full parameter space, $\log \mathcal{I} = -0.59$, both of which are comparable to the discordance between KiDS and Planck in $\Lambda$CDM. 

\begin{figure*}
\begin{center}
\resizebox{8.8cm}{!}{\includegraphics{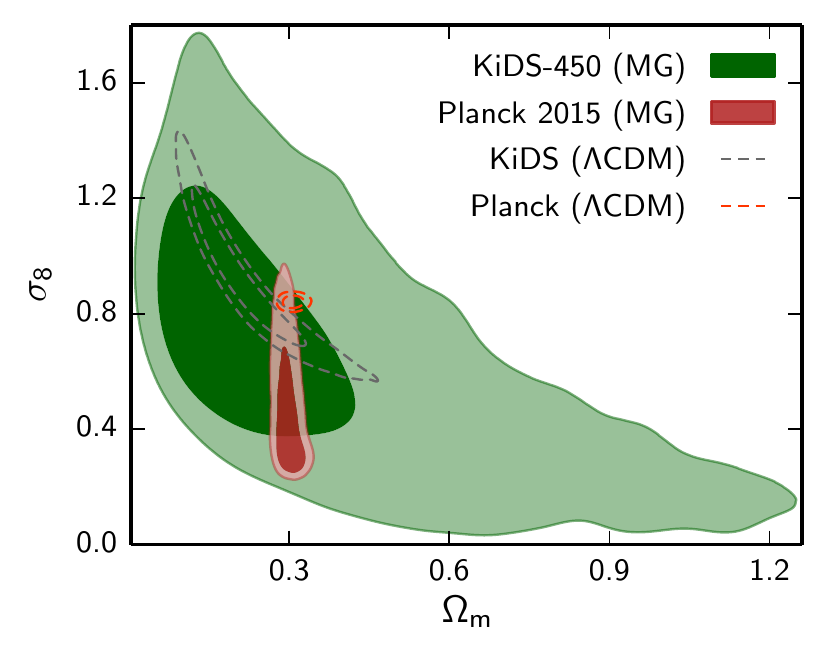}}
\resizebox{8.8cm}{!}{\includegraphics{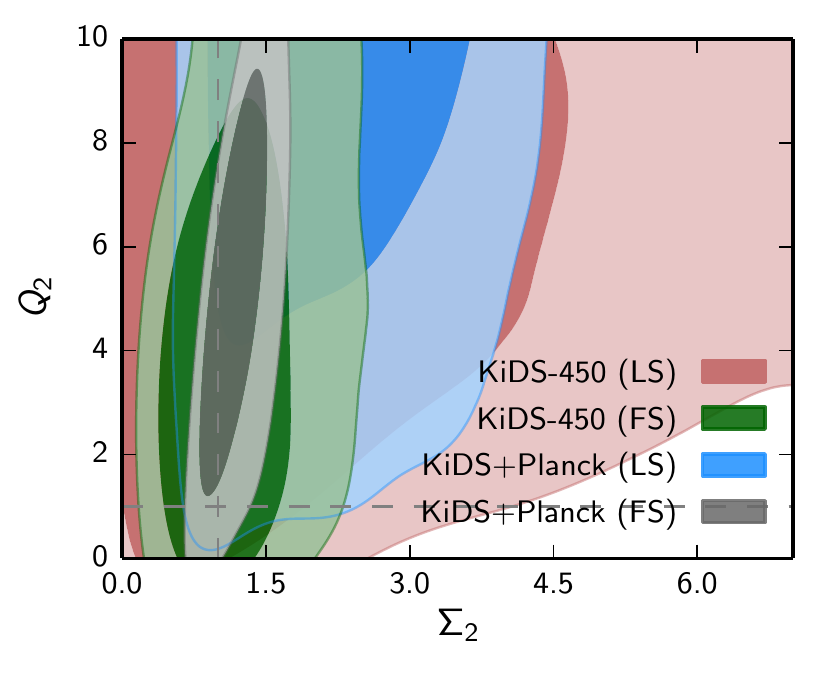}}
\end{center}
\vspace{-2.0em}
\caption{Left: Marginalized posterior contours in the $\sigma_8 - \Omega_{\mathrm m}$ plane (inner 68\%~CL, outer 95\%~CL) in a universe with modified gravity for KiDS in green and Planck in red. For comparison, dashed contours assume fiducial $\Lambda$CDM. Right: Marginalized posterior contours in the $Q_2 - \Sigma_2$ plane for KiDS with fiducial angular scales in green (denoted by `FS'), KiDS keeping only the largest angular scales in pink (denoted by `LS'), and respectively combined with Planck in grey and blue. The indices represent a particular combination of MG bins, such that $z < 1$ and $k > 0.05~h~{\rm{Mpc}}^{-1}$. The dashed lines intersect at the GR prediction ($Q = \Sigma = 1$).
}
\label{figmg}
\end{figure*}

In the $w - \Omega_k$ plane, KiDS agrees with Planck and is concordant with the standard cosmological model, while Planck differs by $\gtrsim2\sigma$ from flat $\Lambda$CDM. 
As the Planck constraint on the dark energy equation of state is weak, this is mainly driven by Planck's propensity to deviate from flatness (similar to that found in Section~\ref{curvsec}).
Weak lensing and the CMB would constitute a powerful combination, but we do not provide joint constraints on the extended degrees of freedom from KiDS and Planck as the two datasets are in tension.
In Figure~\ref{fighubble}, we show the Planck constraint on the Hubble constant in the extended cosmology. 
Due to the severe geometric degeneracy (given the simultaneous consideration of $\Omega_k$ and $w$), the Hubble constant is largely unbounded, with $0.40 < h < 0.91$ at 95\% CL (pushing against the lower end of the prior). 
In Figure~\ref{figampia}, we find that the KiDS constraint on the intrinsic alignment amplitude is degraded to $-0.78 < A_{\rm IA} < 3.4$ (95\% CL), increasingly consistent with no intrinsic alignments. 

When examining the viability of the additional degrees of freedom from the point of model selection, KiDS shows no preference from $\Lambda$CDM (with ${\Delta{\rm DIC}} \approx 0$), while Planck weakly favors the extended cosmological model (with ${\Delta{\rm DIC}} = -3.7$, degraded from $\Delta \chi^2_{\rm eff} = -6.2$ due to the increase in the Bayesian complexity). This weak preference for the extended cosmological model is mainly driven by the nonzero curvature (similar to the result in Section~\ref{curvsec}), and is unlikely to persist with the inclusion of probes that drive the constraint on the curvature towards zero (e.g.~BAOs, \citealt{planck15}).

\begin{figure*}
\begin{center}
\resizebox{8.8cm}{!}{\includegraphics{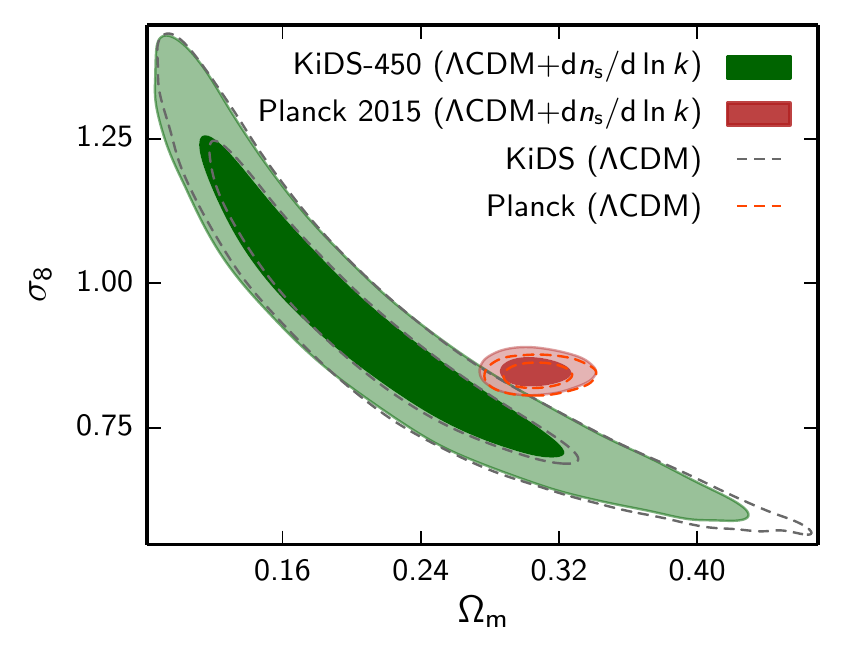}}
\resizebox{8.8cm}{!}{\includegraphics{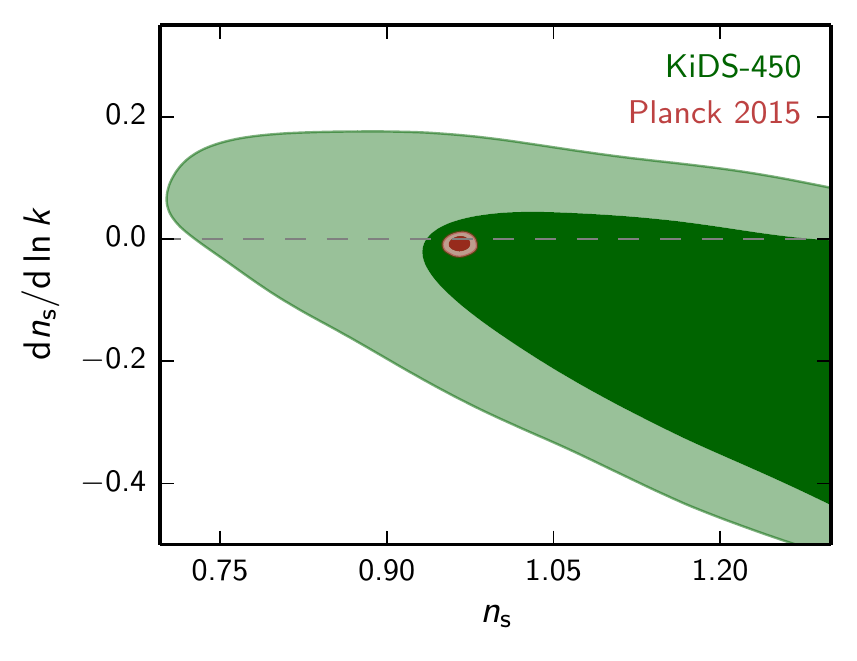}}
\end{center}
\vspace{-2.0em}
\caption{Left: Marginalized posterior contours in the $\sigma_8 - \Omega_{\mathrm m}$ plane (inner 68\%~CL, outer 95\%~CL) in a universe with nonzero running of the scalar spectral index for KiDS in green and Planck in red. For comparison, dashed contours assume fiducial $\Lambda$CDM. Right: Marginalized posterior contours in the ${{\mathrm d}n_{\mathrm s} / {\mathrm d}\ln k} - n_{\mathrm s}$ plane for KiDS in green and Planck in red. The horizontal lines denotes the cosmology with no running of the spectral index.
}
\label{figrun}
\end{figure*}

\subsection{Modified gravity}
\label{modgrav}

We now examine to what extent KiDS and Planck can constrain deviations from General Relativity (GR), and to what extent model-independent modifications to gravity can resolve the relative discordance between these datasets (for model-independent constraints on modified gravity using other data combinations, see e.g.~\citealt{daniel10,johnson15,planckmg15,dVMS16}). To this end, we use \isitgr \citep{dossett11,dossett12}, which is an integrated set of modified modules in \cosmomc designed to test gravity on cosmic scales.

We modify gravity in two ways. Given the first-order perturbed Einstein equations, the first modification takes the form of an effective gravitational constant that enters the Poisson equation:
\begin{equation}
k^2 \phi = -4 \pi G a^2 \sum_i \rho_i \Delta_i Q(k,a) ,
\end{equation}
where $\phi$ is the potential describing spatial perturbations to the metric in the conformal Newtonian gauge, $\rho_i$ is the density of species~$i$, 
$G$ is Newton's gravitational constant, and $Q(k,a)$ encodes the time and scale dependent modifications to the Poisson equation (e.g. \citealt{jz08,bt10,dipd15}, also see \citealt{mb95}). 
The rest-frame overdensity is given by $\Delta_i \equiv \delta_i + 3Ha(1+w_i)\theta_i/k^2$, 
where $\delta_i$ is the fractional overdensity, 
$w_i$ is the equation of state, and $\theta_i$ is the peculiar velocity divergence.
Thus, we can construct an effective gravitational constant, $G_{\rm eff}(k,a) = G \times Q(k,a)$, where $Q \equiv 1$ in GR. 
The second modification to standard gravity enters
\begin{equation}
k^2 [\psi - R(k,a) \phi] = -12 \pi G a^2 \sum_i \rho_i \sigma_i (1+w_i) Q(k,a) ,
\end{equation}
where $\psi$ is the potential describing temporal perturbations to the metric in the conformal Newtonian gauge, 
and $\sigma_i$ is the anisotropic shear stress. Thus, $R(k,a)$ allows the two metric potentials to differ even in the absence of anisotropic stress, and is equivalent to unity in GR. In our MCMC calculations, we substitute $R$ with a parameter that is more directly probed by weak lensing: $\Sigma = Q(1+R)/2$. In general modified gravity (MG) scenarios, the parameters $Q$ and $\Sigma$ can be functions of both scale and time, and affect the growth of structure.

We show the impact of the modified gravity parameters on the shear correlation functions in Figure~\ref{figcorr}, finding that the lensing observables are fairly insensitive to changes in the gravitational constant, while $\Sigma$ effectively boosts or suppresses the observables uniformly across tomographic bin and angular scale unless the parameter possesses time and scale dependence.
In constraining modified gravity, we divide $Q$ and $\Sigma$ in two redshift bins and two scale bins each, with transitions at $k = 0.05~h~{\rm{Mpc}}^{-1}$ and $z = 1$. Thus, $Q_1$ and $\Sigma_1$ correspond to the $\{{\rm low}~z, {\rm low}~k\}$ bins, $Q_2$ and $\Sigma_2$ correspond to the $\{{\rm low}~z, {\rm high}~k\}$ bins, $Q_3$ and $\Sigma_3$ correspond to the $\{{\rm high}~z, {\rm low}~k\}$ bins, $Q_4$ and $\Sigma_4$ correspond to the $\{{\rm high}~z, {\rm high}~k\}$ bins.
This results in 8 MG degrees of freedom varied in our MCMC calculations in addition to the vanilla and lensing systematics parameters (along with the CMB degrees of freedom when applicable). We keep the background expansion to be that of $\Lambda$CDM. In calculating the shear correlation functions, we modify our lensing module to integrate directly over the power spectrum of the sum of the two metric potentials, which in GR reduces to the standard integration over the matter power spectrum.

In Figure~\ref{figmg}, we show constraints in the $\sigma_8 - \Omega_{\mathrm m}$ and $Q_2 - \Sigma_2$ planes, where the indices represent a particular combination of modified gravity bins, such that $z < 1$ and $k > 0.05~h~{\rm{Mpc}}^{-1}$. Since there exists no adequate prescription for the matter power spectrum on nonlinear scales in a cosmology with binned modified gravity (and also no screening mechanism), we consider two distinct cases: one where the fiducial angular scales of KiDS are included (described in Section~\ref{theobs}), and a second case where effectively only linear scales are included in the analysis. For the latter case, we consider the same `large-scale' cuts as in Section~\ref{lcdmsec}, removing all angular scales except for $\theta = \{24.9, 50.7\}$ arcmins in $\xi_{+}^{ij}(\theta)$ and $\theta = 210$ arcmins in $\xi_{-}^{ij}(\theta)$.

For consistency with the previous sections, we show the constraints in the $\sigma_8 - \Omega_{\mathrm m}$ plane for KiDS with fiducial choice of angular scales (presenting the results for KiDS with large-scale cut in Tables~\ref{table:chidic}~and~\ref{table:conc}). The KiDS and Planck contours completely overlap, both as a result of Planck largely losing its ability to constrain $\sigma_8$ for a given matter density, but also because the KiDS constraints are extremely weak given the introduction of eight additional degrees of freedom. 
Thus, the KiDS and Planck $S_8$ constraints agree to within $1\sigma$ (for both choices of scale cuts). As shown in Table~\ref{table:conc}, when accounting for the full parameter space, $\log \mathcal{I} = 0.42$ corresponding to substantial concordance between KiDS and Planck when considering the fiducial angular scales in KiDS, and $\log \mathcal{I} = 1.4$ corresponding to strong concordance between KiDS and Planck when employing large-scale cuts.

Meanwhile, as shown in Figure~\ref{fighubble}, the Planck constraint on the Hubble constant in the extended cosmology moves marginally towards the \citet{riess16} result, where $0.66 < h < 0.71$ (95\% CL), such that the two are still in discordance. In our MG cosmology, the intrinsic alignment amplitude is marginally pushed towards larger values (as compared to the IA amplitude in $\Lambda$CDM) such that the constraint is $-0.039 < A_{\rm IA} < 3.1$ (95\% CL) for KiDS, and $-0.033 < A_{\rm IA} < 2.3$ (95\% CL) for KiDS+Planck. However, the constraints degrade significantly when employing large-scale cuts, such that $-5.2 < A_{\rm IA} < 5.1$ (95\% CL) for KiDS, and $-2.1 < A_{\rm IA} < 2.5$ (95\% CL) for KiDS+Planck. The IA amplitude constraint for KiDS with a large-scale cut in a MG cosmology can be contrasted with the corresponding constraint in $\Lambda$CDM, which at $-5.0 < A_{\rm IA} < 3.2$ (95\% CL) is also fully consistent with zero.

In the $Q_2 - \Sigma_2$ plane, the KiDS constraints are consistent with GR, and mainly sensitive to $\Sigma_2$ as expected. The modified gravity constraints from KiDS are weak for most of the MG parameters, and significantly degraded when keeping only large angular scales, given the significant reduction in the size of the data vector and information contained in the KiDS measurements. The agreement with GR persists when combining KiDS with Planck, not only for $Q_2$ and $\Sigma_2$, but for the other MG parameters as well, 
where the constraints on $\Sigma_i$ are significantly tighter than the constraints on $Q_i$, for both choices of scale cuts (often by an order of magnitude). As shown in Figure~\ref{figmgsub}, the minor exception to the GR agreement is $Q_2 > 2.2$ (at 95\% CL, which reduces to 0.84 at 99\% CL) for KiDS+Planck where a large-scale cut is employed.

Given our particular model of modified gravity, the goodness of fit improves moderately as compared to GR (with $\Delta \chi^2_{\rm eff} \approx -4$ for both KiDS and Planck, and their joint analysis when fiducial angular scales are considered, and by $\Delta \chi^2_{\rm eff} \approx -6$ when large-scale cuts are employed), but this is understandable given the introduction of eight additional degrees of freedom. When examining the difference in DIC between our modified gravity model and GR, we find no preference in favor of modified gravity (with ${\Delta{\rm DIC}} \approx 6$ for Planck, ${\Delta{\rm DIC}} \approx 0$ for KiDS and KiDS+Planck when fiducial scales are considered, ${\Delta{\rm DIC}} \approx 6$ for KiDS with a large-scale cut, and ${\Delta{\rm DIC}} \approx 2$ for KiDS+Planck with a large-scale cut). A next step would be to consider more model-dependent approaches to constraining modified gravity, but we leave further investigations~of these models and their potential viability to forthcoming analyses.

\subsection{Running of the spectral index}
\label{runningsec}

Lastly, beyond the curvature of the universe, we also relax the strong inflation prior on the running of the scalar spectral index, ${{\mathrm d}n_{\mathrm s} / {\mathrm d}\ln k}$, defined via the 
dimensionless power spectrum of primordial curvature perturbations,
\begin{equation}
\ln P_{\mathrm s} (k) = \ln A_{\mathrm s} + (n_{\mathrm s} - 1) \ln\left({k \over k_{\rm pivot}}\right) + {1\over2} {{\mathrm d}n_{\mathrm s} \over {\mathrm d}\ln k} \ln\left({k \over k_{\rm pivot}}\right)^2,
\label{eq:plaw}
\end{equation}
where $A_{\mathrm s}$, $n_{\mathrm s}$, and ${{\mathrm d}n_{\mathrm s} / {\mathrm d}\ln k}$ are evaluated at the pivot wavenumber $k_{\rm pivot}$ listed in Table~\ref{table:priors}. 
While most popular inflation models predict $\left|{{\mathrm d}n_{\mathrm s} / {\mathrm d}\ln k}\right| \lesssim 10^{-3}$ \citep{kt95}, large negative running can be generated by multiple fields, temporary breakdown of slow-roll, or several distinct inflationary stages (e.g.~\citealt{baumann08} and references therein).

In Figure~\ref{figcorr}, we show the imprint of a nonzero running of the scalar spectral index on the lensing observables. As expected, through its impact on the matter power spectrum, a negative running provides a scale-dependent suppression of the shear correlation functions that increases towards small angular scales, and is particularly correlated with the imprint of baryon feedback.
We show the resulting constraints in the $\sigma_8 - \Omega_{\mathrm m}$ and ${{\mathrm d}n_{\mathrm s} / {\mathrm d}\ln k} - n_{\mathrm s}$ planes in Figure~\ref{figrun}. In the $\sigma_8 - \Omega_{\mathrm m}$ plane, it is evident that the introduction of nonzero running does not alleviate the tension between KiDS and Planck, with the respective contours only marginally affected by the extended degree of freedom. Analogous to the $\Lambda$CDM results, the tension in the $S_8$ parameter is at the $2.3\sigma$ level, and $\log \mathcal{I} = -0.66$ corresponding to `substantial discordance' between the KiDS and Planck datasets. 

When examining the constraints in the ${{\mathrm d}n_{\mathrm s} / {\mathrm d}\ln k} - n_{\mathrm s}$ plane, we find weak constraints on both parameters from KiDS. However, KiDS does independently from Planck agree with zero running of the scalar spectral index (marginalized constraint of $-0.40 < {{\mathrm d}n_{\mathrm s} / {\mathrm d}\ln k} < 0.15$ at 95\% CL). As expected, the Planck constraint on the running is substantially more competitive, and would require significantly more precise lensing measurements to improve. 
Meanwhile, in the extended cosmology, the Planck constraint on the Hubble constant and the KiDS constraint on the intrinsic alignment amplitude are both close to the respective constraints in $\Lambda$CDM.
The extended cosmology does not improve the goodness of fit noticeably as compared to $\Lambda$CDM (with $\Delta \chi^2_{\rm eff} \approx -1$ for KiDS and and $\Delta \chi^2_{\rm eff} \approx 0$ for Planck), and is not favored by the KiDS and Planck datasets (with ${\Delta{\rm DIC}} \lesssim 1$ for KiDS and Planck).

\section{Conclusions}
\label{conclusions}

We have performed an extended lensing systematics and cosmology analysis of the tomographic weak gravitational lensing measurements of the Kilo Degree Survey (KiDS; \citealt{dejong13, kuijken15, Hildebrandt16}). 
The extended lensing systematics include non-informative priors on 
the amplitude and redshift-dependence of intrinsic galaxy alignments, and 
baryonic feedback modifying the nonlinear matter power spectrum. 
In Appendix A, we further explore the impact of increasing our uncertainty on either the shear calibration correction, or the photometric redshift distributions, or indeed any systematic that changes the amplitude of the weak lensing signal.
Meanwhile, the extended cosmologies with fiducial treatment of the systematic uncertainties include massive neutrinos, nonzero curvature, evolving dark energy, modified gravity, and running of the spectral index. The aim of this paper has been three-fold. We have examined to what extent the extended models can be constrained by KiDS, to what extent they are favored as compared to the standard cosmological model, and to what extent they can alleviate the discordance between KiDS and Planck CMB temperature measurements.

To this end, we use the same KiDS measurements, fitting pipeline, and approach to systematic uncertainties as in \citet{Hildebrandt16}. 
In addition to the standard $\Lambda$CDM parameters, 
we always vary the intrinsic alignment and baryon feedback amplitudes (fiducially with informative priors).
We do not vary any parameters in our treatment of the photometric redshift uncertainties, but instead capture the uncertainties with 1000 bootstrap realizations of the tomographic redshift distributions.
Unlike \citet{Hildebrandt16}, we do not fiducially impose an informative prior on the Hubble constant from \citet{riess16}, which extends our contours along the lensing degeneracy direction but does not particularly affect the discordance with Planck. 

In a $\Lambda$CDM cosmology with fiducial treatment of lensing systematics, the discordance between KiDS and Planck is $2.1\sigma$ in $S_8 = \sigma_8 (\Omega_{\mathrm m}/0.3)^{0.5}$.
In evaluating the level of discordance over the full parameter space, we use the $\log \mathcal{I}$ statistic grounded in information theory. Similar to the result in \citet{Hildebrandt16}, we find $\log \mathcal{I} = -0.63$, which corresponds to `substantial discordance' between the two datasets.
As we move beyond the fiducial model, our findings are summarized below:

\begin{enumerate}
\item 
{\it Extended lensing systematics}: 
We impose non-informative
priors on the intrinsic alignment and baryon feedback amplitudes ($A_{\rm IA}$ and $B$), and 
introduce $\eta_{\rm IA}$ that governs the redshift dependence of the intrinsic alignment signal. 
These parameters are constrained to $B < 4.6$ (95\% CL), $-0.45 < A_{\rm IA} < 2.3$ (95\% CL), and $-16 < \eta_{\rm IA} < 4.7$ (95\% CL). The constraints are consistent with the fiducial treatment of lensing systematics, and do not particularly affect the discordance between KiDS and Planck. The discordance between the datasets remains even when removing the smallest angular scales in KiDS most sensitive to nonlinear physics, or allowing for a large uncertainty in the amplitudes of the shear correlation functions bin due to unknown systematics. As we step through each of the extended cosmologies below, 
the KiDS constraint on the intrinsic alignment amplitude is remarkably robust with a consistent $2\sigma$ positive deviation from zero.

\item {\it Neutrino mass}: 
We capture the effects of neutrino mass on the nonlinear matter power spectrum with an updated version of 
$\hmcode$ \citep{Mead16}. KiDS constrains $\sum m_{\nu} < 4.0$ eV (95\% CL), which does not bring about concordance between KiDS and Planck, and is not required by the data.

\item {\it Curvature}: 
KiDS and Planck independently constrain the curvature to be positive at about 95\%~CL. 
Employing model selection criteria, nonzero curvature is not favored by KiDS, and weakly favored by Planck.
The extra degree of freedom increases the discordance between the datasets to $3.5\sigma$ in $S_8$, and to $\log \mathcal{I} = -1.7$ (corresponding to `strong discordance'). 

\item {\it Dark energy (constant $w$)}: 
A constant dark energy equation of state $w$ brings `substantial-to-strong' concordance between KiDS and Planck. In this cosmology, the Planck constraint on the Hubble constant is wider and in agreement with \citet{riess16}. KiDS and Planck are separately in agreement with a cosmological constant, but the combined analysis of KiDS and Planck with a uniform prior on $H_0$ from \citet{riess16} deviates by $2\sigma$ from $w = -1$. From the point of model selection, the extended model is weakly favored as compared to $\Lambda$CDM.

\item {\it Dark energy ($w_0-w_a$)} 
A time-dependent parameterization of the dark energy equation of state brings substantial concordance between KiDS and Planck, and removes the $H_0$ tension between Planck and \citet{riess16}. KiDS is in agreement with a cosmological constant, while Planck shows a $2\sigma$ deviation. Combining KiDS and Planck with a uniform $H_0$ prior from \citet{riess16} gives a $3\sigma$ deviation from a cosmological constant that is moderately favored by the data. This deviation from a cosmological constant is consistent with SN distance measurements from the `Joint Light-curve Analysis' sample (JLA; \citealt{Betoule13, Betoule14}), 
but in tension with BAO measurements from the 6dF Galaxy Survey \citep{beutler11}, SDSS Main Galaxy Sample \citep{ross15}, and BOSS LOWZ/CMASS samples \citep{anderson14} when combined with Planck. 
Meanwhile, the BAO+Planck constraints are separately in tension with \citet{riess16}. The constraints from KiDS+JLA and KiDS+BAO are weaker, in agreement both with KiDS+Planck+$H_0$ and with a cosmological constant.

\item {\it Curvature + dark energy (constant $w$)}: 
Beyond unitary extensions to the underlying cosmology, we simultaneously vary $\Omega_k$ and $w$. 
The impacts of the two degrees of freedom partially cancel, such that the discordance between KiDS and Planck is similar to that in $\Lambda$CDM. The extra degrees of freedom are only weakly constrained by KiDS, and not favored by the data.

\item {\it Modified gravity}: 
Introducing parameters that govern deviations to the Poisson equation $Q(k,a)$ and deflection of light $\Sigma(k,a)$, divided in two redshift bins and two scale bins each, removes the discordance between KiDS and Planck. However, the extra degrees of freedom are not favored by the data, and the MG constraints 
are in agreement with GR.

\item {\it Running of the spectral index}: 
The KiDS/Planck discordance is only marginally affected by a running of the spectral index. 
Independently from other probes, KiDS constrains the running to be consistent with zero ($-0.40 < {{\mathrm d}n_{\mathrm s} / {\mathrm d}\ln k} < 0.15$ at 95\% CL). 
\end{enumerate}

To conclude, the discordance between KiDS and Planck is largely robust to changes in the lensing systematics and underlying cosmology. The most interesting exception to this is a cosmology with a time-dependent dark energy equation of state, which provides substantial concordance between KiDS and Planck, 
is $3\sigma$ discrepant from the cosmological constant scenario, and moderately favored by KIDS+Planck. 
The KiDS data are publicly available at \kidsaddress. 
We also make the fitting pipeline and data that were used in this analysis public at \sjaddress. 

\section*{Acknowledgements}
We much appreciate useful discussions with Alexandre Barreira, Jason Dossett, Manoj Kaplinghat, Antony Lewis, Nikhil Padmanabhan, David Parkinson, and Martin White. We thank Simon Forsayeth, Robin Humble, and Jarrod Hurley for HPC support. 
We also thank George Efstathiou for useful discussions about internal consistency tests, and the anonymous referee for their helpful comments on this paper.
We acknowledge the use of \astac and \caastro time on Swinburne's swinSTAR and NCI's Raijin machines. We acknowledge the use of \camb and \cosmomc packages (\citealt{Lewis:2002ah}; \citealt{LCL}). 
This work is based on data products from observations made with ESO Telescopes at the La Silla Paranal Observatory under programme IDs 177.A-3016, 177.A-3017 and 177.A-3018.
Parts of this research were conducted by the Australian Research Council Centre of Excellence for All-sky Astrophysics (CAASTRO), through project number CE110001020.  This work was supported by the Flagship Allocation Scheme of the NCI National Facility at the ANU.
This work was performed in part at the Aspen Center for Physics, which is supported by National Science Foundation grant PHY-1066293.
AM acknowledges support from a CITA National Fellowship.
CB acknowledges the support of the Australian Research Council through the award of a Future Fellowship.
AC acknowledges support from the European Research Council under the FP7 grant number 240185.
JdJ is supported by the Netherlands Organisation for Scientific Research (NWO) through grant 614.061.610.
IFC acknowledges the use of computational facilities procured through the European Regional Development Fund, Project ERDF-080 `A supercomputing laboratory for the University of Malta'.
CH acknowledges support from the European Research Council under grant numbers 240185 and 647112.
HHi is supported by an Emmy Noether grant (No. Hi 1495/2-1) of the Deutsche Forschungsgemeinschaft.
HHo acknowledges support from the European Research Council under FP7 grant number 279396.
BJ acknowledges support by an STFC Ernest Rutherford Fellowship, grant reference ST/J004421/1.
DK and PS are supported by the Deutsche Forschungsgemeinschaft in the framework of the TR33 `The Dark Universe'.
FK acknowledges support from a de Sitter Fellowship of the Netherlands Organization for Scientific Research (NWO) under grant number 022.003.013.
KK acknowledges support by the Alexander von Humboldt Foundation.
LM is supported by STFC grant ST/N000919/1.
MV acknowledges support from the European Research Council under FP7 grant number 279396 and the Netherlands Organisation for Scientific Research (NWO) through grant 614.001.103.

\bibliographystyle{mn2e}
\bibliography{extendedkids}

\appendix

\section{Impact of Unknown Systematics}
\label{sublab2}

\begin{figure*}
\includegraphics[width=0.97\hsize]
{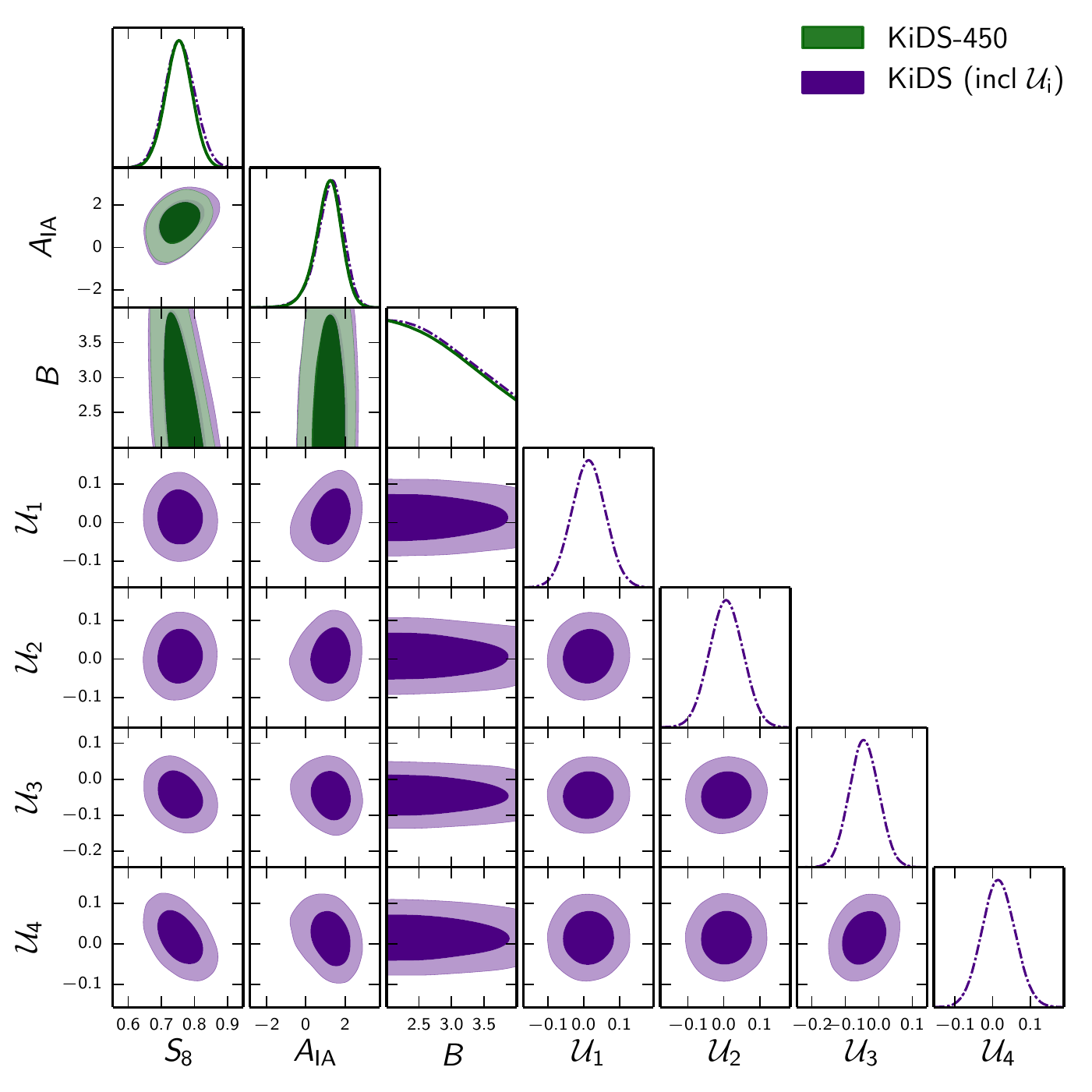}
\vspace{-1.3em}
\caption{\label{figerrsub} Posterior distributions of the $S_8 = \sigma_8 \sqrt{\Omega_{\mathrm m}/0.3}$ parameter combination, intrinsic alignment amplitude $A_{\rm IA}$, baryon feedback $B$, unknown sources of systematic amplitudes $\mathcal{U}_i$, and their correlation. 
The constraints in green (solid) correspond to the fiducial KiDS analysis, where $\mathcal{U}_i = 0$, while the constraints in purple vary the $\mathcal{U}_i \in (-0.3, 0.3)$ with Gaussian priors of $\sigma(\mathcal{U}_i) = 0.05$ along with the other parameters.
The priors on other parameters are listed in Table~\ref{table:priors}. In this figure, we do not show the primary $\Lambda$CDM parameters that were simultaneously varied in the MCMC.
}
\end{figure*}

As with all scientific analyses we cannot categorically rule out that there are additional unknown sources of systematic uncertainties that have not been considered in our analysis (take for example the `GI' intrinsic alignment term which is now considered, but was unknown to the weak lensing community until \citealt{HS04}). 
To explore the impact of increasing our uncertainty on either the shear calibration correction, or the photometric redshift distributions, or indeed any systematic that changes the amplitude of the weak lensing signal, we show in Figure~\ref{figerrsub} the submatrix of constraints on the amplitudes $\mathcal{U}_i$ in each of the four tomographic bins such that $\xi_{\pm}^{ij}(\theta) \rightarrow (1+\mathcal{U}_i)(1+\mathcal{U}_j) \xi_{\pm}^{ij}(\theta)$ with Gaussian priors arbitrarily chosen to have a width $\sigma(\mathcal{U}_i) = 0.05$. These additional nuisance parameters can be compared to the constraints on the intrinsic alignment amplitude, the baryonic feedback parameter, and the derived $S_8 = \sigma_8 \sqrt{\Omega_{\mathrm m}/0.3}$ parameter. We do not show constraints on the primary $\Lambda$CDM parameters, which are simultaneously varied in the analysis. 
Despite the wide priors on the $\mathcal{U}_i$ parameters, there is only a 15\% increase in the uncertainty on $S_8$. We find $S_8 = 0.756 \pm 0.046$ in the extended analysis as compared to $S_8 = 0.752 \pm 0.040$ in the fiducial analysis. In this, rather arbitrary, case the discordance with Planck would decrease by $0.3\sigma$ (such that the tension is still at the $2\sigma$ level).

This test both verifies the Fisher matrix analysis in Appendix~A of \citet{Hildebrandt16}, and allows us to look for internal consistency between the different tomographic slices. We find that the constraints on $\mathcal{U}_i$ are dominated by the prior, with the posterior means all consistent with zero such that the tomographic slices are consistent with each other. The largest amplitude shift can be seen in the third tomographic bin where the lensing measurements are comparably lower than the other tomographic bins. The likely cause of this slight amplitude change is the presence of small-angular scale, low amplitude B-modes that predominantly affect the third tomographic bin \citep{Hildebrandt16}.

We find that the fit to the data does not particularly improve when including these four additional degrees of freedom, and the change in $\Delta{\rm{DiC}} \approx 5$, such that this extended unknown systematics model is not favored by the data.

\section{Modified gravity subspace}
\label{sublab}

\begin{figure*}
\includegraphics[width=0.97\hsize]
{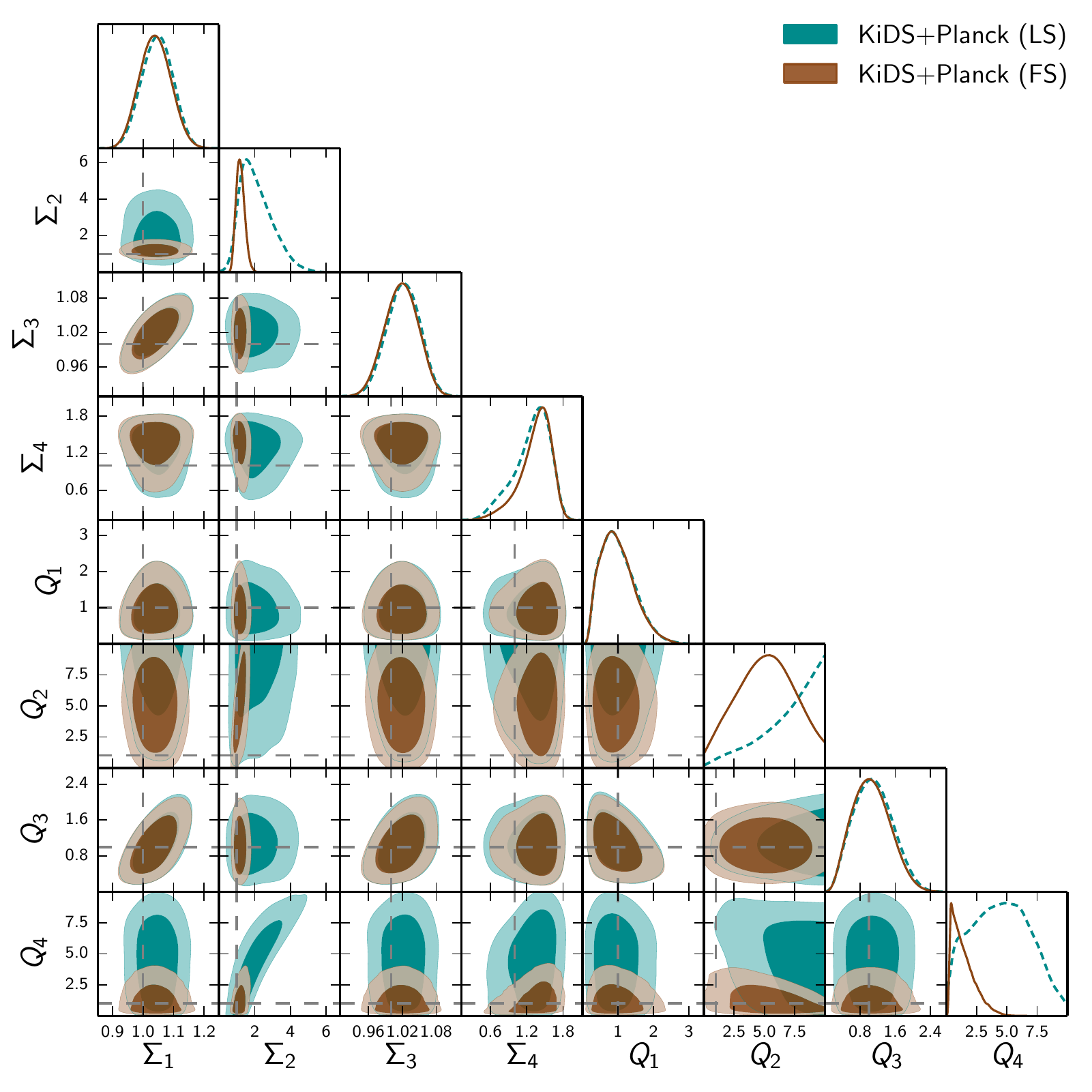}
\vspace{-1.3em}
\caption{\label{figmgsub} Posterior distributions of the modified gravity parameters and their correlation. 
The constraints in brown (solid) correspond to KiDS+Planck, considering the fiducial angular scales of KiDS (listed in Section~\ref{theobs}), while the constraints in cyan (dashed) correspond to KiDS+Planck keeping only the largest, effectively linear, scales of KiDS (Section~\ref{modgrav}).
Parameter definitions and priors are listed in Table~\ref{table:priors}. The bin transitions are at $k = 0.05~h~{\rm{Mpc}}^{-1}$ and $z = 1$. The indices are devised such that 
$Q_1$ and $\Sigma_1$ correspond to the $\{{\rm low}~z, {\rm low}~k\}$ bins, $Q_2$ and $\Sigma_2$ correspond to the $\{{\rm low}~z, {\rm high}~k\}$ bins, $Q_3$ and $\Sigma_3$ correspond to the $\{{\rm high}~z, {\rm low}~k\}$ bins, $Q_4$ and $\Sigma_4$ correspond to the $\{{\rm high}~z, {\rm high}~k\}$ bins. GR is given by $Q = \Sigma = 1$. In this figure, we do not show the other fiducial lensing and CMB parameters that were simultaneously varied in the MCMC.
}
\end{figure*}

In Figure~\ref{figmgsub}, we show the submatrix of binned modified gravity constraints obtained in the analysis presented in Section~\ref{modgrav}. 

\end{document}